\documentclass[times,12pt]{article}
\pdfoutput=1
\usepackage{amsmath,amssymb,amsfonts,latexsym,amsthm,enumerate,url}
\usepackage{latex8}
\usepackage{times}
\usepackage{graphicx}
\usepackage{mathrsfs}
\date{}

\newtheorem{thm}{Theorem}[section]

\newtheorem{proposition}[thm]{Proposition}

\newcommand{\beq}[1]{\begin{equation}\label{#1}}
\newcommand{\enq}[0]{\end{equation}}

\newcommand{\remove}[1]{}

\newcommand{\comment}[1]{}

\title{Statistical Aspects of the Quantum Supremacy Demonstration}
\author {Yosef Rinott\\  The Hebrew University of Jerusalem \\ Federmann Center for the Study of Rationality and Department of Statistics
        \and Tomer Shoham\\ The Hebrew University of Jerusalem\\ Federmann Center for the Study of Rationality  and Department of Statistics
\and    Gil Kalai\\ The Hebrew University of Jerusalem\\
  Einstein Institute of Mathematics and\\Efi Arazi School of Computer Science, IDC,
  Herzliya.\thanks{Federmann Center for the Study of Rationality, and Quantum Information Science Center at HUJI, FACT center at IDC.
    Supported by ERC advanced grant 834735.}
}
\begin{document}

    {    \maketitle}

\begin{abstract}
	In quantum computing, a demonstration of \textit{quantum supremacy} (or quantum advantage) consists of presenting a task, possibly of no practical value, whose computation is feasible on a quantum device, but   cannot be performed by classical computers  in any feasible amount of time. 
The notable claim of quantum supremacy
presented by Google's team in 2019 consists of demonstrating the ability of a quantum circuit to generate, albeit with considerable noise, bitstrings
from a distribution that is
considered hard to simulate on classical computers. 
Very recently, in 2020, a quantum supremacy claim was presented by a group from the University of Science and Technology of China, using a different technology and generating a different distribution,  but sharing some statistical  principles with Google's demonstration.  

Verifying that the
generated data is indeed from the claimed distribution and assessing the circuit's noise level and its fidelity is a
statistical undertaking.
The objective of this paper is to explain the relations between
quantum computing and some of the statistical aspects
involved in demonstrating  quantum supremacy
in terms that are accessible to statisticians, computer scientists, and mathematicians.
Starting with the  statistical modeling and analysis in Google's demonstration, which we explain, we study various   estimators of the fidelity, and  different approaches to testing the distributions generated by the quantum computer.   We propose different noise models, and discuss their implications. A preliminary study of the Google data, focusing mostly on circuits of 12 and 14 qubits is given in different parts of the paper. 
\end{abstract}

%\begin{keyword}
%\kwd{Google's quantum computer}
%\kwd{random distributions}
%\kwd{estimation of sampling weights}
%\kwd{size bias}
%\kwd{\LaTeXe}
%\end{keyword}

%\end{frontmatter}

	\section{Introduction}	\label{sec:Int}
Google's announcement of  quantum supremacy \cite{Google} was compared by
various writers to landmark achievements such as the Wright brothers'
invention of a motor-operated airplane, Fermi's demonstration of a nuclear chain reaction, and the discovery of the Higgs boson.
It was also met with some skepticism, which is partly due to  the fact that Google's quantum
computer, and quantum computers in general, have not at this point performed any practical task (such as factoring large integers). 
Instead, Google's quantum
computer performs a \textit{sampling task}; that is,
it generates random bitstrings,  with considerable noise, from a discrete distribution supported on
$M$ values, with probabilities whose computations are far beyond the reach of classical computers
for large $M$. (Therefore, classical computers cannot carry out this sampling task.)
A quantum computer, suitably adjusted, is
capable of sampling from such a distribution without having to compute the probabilities
explicitly, and thus could claim quantum supremacy.
Google's team estimated that the task performed by their quantum circuit would take 10,000 years on classical computer. In a recent quantum advantage demonstration, see \cite{Hefei}, a team from the University of Science and Technology of China (USTC) claimed that the sampling task computed by their quantum computer, which differs from Google's, would take 2.5 billion years to perform on a classical supercomputer! \footnote{However, the method of \cite{KK} may lead to an efficient classical sampling algorithm for this task.}  Both Google's and USTC's quantum computers required about 200 seconds for their tasks. 
 
Verifying that the generated data indeed have the claimed distribution in spite of much
noise is a statistical problem, which we explain and generalize, offering some new statistical aspects. We focus on the statistical aspects of Google's experiment without aiming to pass judgment on its merits. Parts of the discussion below are quite general and could shed some light on other existing or future  noisy intermediate-scale quantum computer experiments.  Aiming for a general mathematical audience, 
we briefly review some basic relevant statistical ideas  and some of the basics of quantum computing,   
and therefore parts of the paper are of an expository nature. Also,   
our notation is a compromise between Google's notation  \cite{Google} and more standard statistical notation.
We start with the main statistical models and problems; 
background on quantum computing is given in Section \ref{QC}.

\subsection{The main model and statistical problem}	

We provide a brief schematic view of the nature of quantum computers by describing Google's quantum computer. More details are given in Section 2.

 A quantum computer (or circuit) consists of $n$ qubits, which are its basic memory units. 
The computer operates via \textit{gates} operating on one and two qubits. In Google's case
qubits are realized as coupled (linked) chips that become superconducting at cold temperatures, resulting in reduced noise and dissipation. A qubit can exist at two excitation levels, that is, energy levels that depend on the orbitals of electrons in the chip, and thus when read/measured, each qubit takes the value 0 or 1 and the whole system yields an $n$-vector of 0's and 1's to which we refer as \textit{bitstring}. Before being measured the system exists in a \textit{superposition} of all $2^n$ $n$-vectors, which
means that all $2^n$ $n$-vectors are represented potentially in the system, and only
when \textit{measured} (observed) it ``collapses" to some $n$-vector, chosen with a certain
probability.
The system operates with a great deal of noise due to
 errors in the qubits and the gates, and to \textit{readout errors} (of the qubits) at the end of the process, affecting the distribution of the observed vectors.  When the circuit is set, by setting its gates, it determines  a set of $M:=2^n$ probabilities $\{w_{{\bf x}^{(i)}} : {\bf x}^{(i)} \in \{0,1\}^n\}$ such that when the system is  measured, it will yield the vector ${\bf x}^{(i)}$ with probability   $w_{{\bf x}^{(i)}}$ for $i=1,\ldots,M$, where $\sum_{i=1}^{M} w_{{\bf x}^{(i)}}=1$, provided it runs without errors. (We may denote $w_{{\bf x}^{(i)}}$ by $w_i$, and  ${\bf x}^{(i)}$  by ${\bf x}_{w_i}$.) If we run the system and measure it $N$ times,  we   get $N$ such $n$-vectors, and denoting the probability of no error, known as the  \textit{fidelity}, by $\phi$, our sample ${\cal S}_{\bf x}$ will consist of $N$ iid vectors  ${\bf x}^{(i)}$,  sampled with probability denoted by $\pi({\bf x}^{(i)})$ or $\pi(w_i)$ satisfying 
\begin{equation}
\label{eq:googsbsample}
\pi({\bf x}^{(i)}) = \pi({ w}_i)=\phi {w_{i}}+(1-\phi)/2^n;
\end{equation}
that is, the vector ${\bf x}^{(i)}$ is chosen with probability $w_i$ in the event that no error occurs (whose  probability is $\phi$), and if an error occurs (with probability $1-\phi$), the vector is chosen uniformly. This is the sampling model proposed in \cite{Google}. It describes the  sampling task distribution  as a mixture of the desired distribution $w_i$ and a uniform distribution on the space of $\{0,1\}^n$. We shall   discuss this model later, along with generalizations of the error model. We denote  the $N$ sampled values by ${\cal S}_{\bf x}=\{\widetilde{\bf x}^{(j)}\}=\{{\bf x}_{\widetilde {w}_j}\}$, and by ${\cal S}_w$  the sample $\{{\widetilde {w}_j}\}$ of the probabilities associated with the sampled vectors in ${\cal S}_{\bf x}$. Note that ${\cal S}_{\bf x}$ and ${\cal S}_w$ are \textit{multisets}, allowing multiple instances of elements as in iid sampling or sampling with replacement. 

For Google's quantum computer one can assume that the probabilities $w_i$ of a randomly chosen quantum circuit (to be discussed later) are random variables generated as follows. Let $z_i$ be iid Exponential(1) variables (with density $e^{-z}$ for $z>0$), $i=1,\ldots,M=2^n$,
and 
$w_{i}={z_{i}}/{\sum_{j=1}^M z_{j}}$. It is well known that 
$(w_1,\ldots,w_M)$ has the Dirichlet($\boldsymbol \alpha$) distribution  with parameter $\boldsymbol \alpha=\textbf{1}$, an $M$-vector  whose components are all equal to 1. This is a uniform distribution over the $(M-1)$-dimensional standard simplex. 
For most purposes 
only a few moments of the
distribution will be taken into account, and our study applies more generally.
Clearly $Ew_i=1/M $, and more generally,       we shall need the following facts: 
\begin{equation}
\label{eq:mom}
 Ew_i^k=k!/[M\cdots(M+k-1)], \, E(w_iw_j)=1/[M(M+1)] \text{  for  } i\neq j;
\end{equation} 
see, e.g., \cite{BJK}.
 The distribution of $\{{\widetilde {w}_j}\}$ differs from that of $\{{{w}_j}\}$. Under the sampling model  \eqref{eq:googsbsample}, with probability $\phi$,  ${w}_j$ is sampled with a probability proportional to ${w}_j$. Such sampling is known as \textit{size-biased sampling}; see  Section \ref{sec:BSZ}.

In \cite{Google} the
sample size is denoted  by $N_s$, which we abbreviate to $N$;  
Google's notation for the  fidelity parameter is $F$, which we denote by $\phi$, and their notation for the random probabilities $w_i$ is ${\cal P}({\bf x}^{(i)})$,
known as a Porter--Thomas distribution. We sometimes use the notation ${\cal P}_C({\bf x}^{(i)})$ to emphasize the dependence of the probabilities  on the (random) circuit $C$.

The statistical problems that arise in relation to the model of \eqref{eq:googsbsample} and its extensions described below include estimating of the parameter $\phi$ and verifying that $\phi>0$, as $\phi=0$ indicates that the circuit produces pure noise, and testing the validity of the model \eqref{eq:googsbsample} and its variations under different assumptions.  

When the $w$'s are considered non-random, possibly by conditioning, this can be seen as sampling from a discrete (finite) population. 
Recalling that the $w$'s are generated randomly
from a given distribution, we say that the sample is generated from a random
discrete distribution in the sense of Kingman \cite{Kingman}. See Section \ref{sec:BSZ} for further details.
The estimation takes advantage of both the process  generating the $w$'s, and the special
nature of the sampling scheme \eqref{eq:googsbsample} or variants of it.

We consider two kinds of analyses of estimators. First, we condition on $\{w_i\}_{i=1}^M$, which amounts to considering a particular quantum circuit, and sampling from the fixed set $\{w_i\}_1^M$.  We study conditional  properties of different estimators, such as their bias and variance.  Second, assuming that the $w_i$'s are random and satisfy  some moment conditions, we study properties of estimators when averaged  over the randomness of $\{w_i\}_1^M$. We compare the two analyses and discuss them in Section \ref{sec:estfi}. 
Google's estimator for the fidelity is quite simple and given by
\begin {equation}
\label {google-estimator-i}
U:= 2^n\frac {1}{N}\sum_{j=1}^N {\cal P}_C(\widetilde{\bf x}^{(j)})\,-1.
\end {equation}
The estimator  $U$
is nearly  unbiased (see \eqref{eq:EUUU}) when both the sample $\{\widetilde{\bf x}^{(j)}\}$ and the
probabilities $w_i={\cal P}({\bf x}^{(i)})$
are random and expectation is taken over both. However, when considering fixed probabilities $w_i$, the estimator $U$ is not unbiased.

In Sections \ref{sec:Gphi} and \ref{sec:condMLE}, respectively, we discuss an unbiased version $V$ of $U$, and the \textit{maximum likelihood estimator} (MLE), which is nearly unbiased. Both turn out to be superior to the above $U$ in terms of  variance and bias for both types of experiments: sampling repeatedly from a single circuit, and averaging over  several circuits. This superiority decreases when $\phi$ is small and $M$ is large. In particular, the improvement achieved by the estimators $U$ and $V$ is insignificant when the number of qubits is above 30. Nevertheless, it does  matter for many current relatively small-scale circuit experiments and to Google's demonstration where small circuits are used for extrapolation arguments to larger circuits; see  \cite {Google} Figure 4, which starts with $n=12$.   
Our statistical  study involves theoretical arguments, demonstrated by simulated data and Google's experimental data.

The present statistical setup brings to mind super-population models (see, e.g., \cite{SSW} Ch. 14.5 or \cite{Nathan}) where
a population $\mathscr{P}=\{w_i\}$ of size $M$ of some  measurements
is considered to be a realization from a continuous or discrete
distribution known as a super-population model,
and then a sample of size $N$ is taken from $\mathscr{P}$, using a known sampling scheme.
A standard goal in this case is to make an inference on the parameters of
the population $\mathscr{P}$ using the sample. 
However, in our case the sampling scheme \eqref{eq:googsbsample} is unknown  and our goal is different; instead of 
estimating the population parameters, we want to estimate the
parameter $\phi$, which is part of the sampling scheme.

We add some more details on the connection of the statistical models described above to quantum physics and  the Google experiment.
(Section \ref {QC} gives a detailed explanation.)
The (ideal) quantum state of a quantum computer with $n$ qubits
is represented by a unit vector ${\bf u}= (u_1,u_2,\dots,u_M)$ in an $M$-dimensional complex vector space, $M=2^n$.
The coordinates of ${\bf u}$ are referred to as amplitudes. We cannot probe these amplitudes directly (this follows from Heisenberg's uncertainty principle),
but we can  measure the state, and 
this yields a single sample from a discrete probability distribution with probabilities $w_1=|u_1|^2,w_2=|u_2|^2,\dots$, $w_M=|u_M|^2$.

Next, come random circuits: when the collection of gates is chosen at random then the vector ${\bf u}$ behaves
like a random unit vector and this implies, when $M$ is large,  that
the random  probabilities  $w_1,\ldots,w_M$
are modeled to arise from exponential $z_i$'s normalized by their sum as above. To see this note first that a
uniformly distributed $M$-dimensional vector on a sphere of radius 1 can be generated by taking $M$ iid N(0,1) variables,
and normalizing the length to 1, which is obtained approximately (for large $M$)  by dividing by $\sqrt{M}$. Therefore, 
if we consider ${\bf u}$ as a unit vector in a real $2M$-dimensional space, the coordinates  of ${\bf u}$ behave like 
iid Gaussian variables, and therefore each squared absolute value of a complex coordinate behaves like the sum of the squares of two
iid Gaussians, which has an exponential distribution. (The sum of the squares of $k$ iid Gaussians is distributed as $\chi^2(k)$, that is,
as a $\chi^2$-distribution with $k$ degrees of freedom. Thus the sum of the squares of two Gaussians has a 
$\chi^2(2)$ distribution, which coincides with a constant times Exp(1).) 

The quantum computer samples $N$ of these $w_i$'s independently according to a model such as \eqref{eq:googsbsample}.
This is the  sampling task.
Given a quantum circuit, the computation of the linked probabilities ${w_i}$ (even a single one of them) can only be done with exponentially increasing efforts
by classical computers if $n< 40$ or so,
and it becomes a practically impossible task if, say, $n>50$ \footnote{Here we refer to the most complicated circuits
	in the Google experiment. There are various simplified circuits for which computing the linked probabilities is feasible.}
(Google's ultimate experiment is with $n=53$). As  classical computers cannot
compute these probabilities for large $n$, they cannot produce samples according to them. 
The quantum computer does not compute these probabilities, and the claimed  supremacy is in its ability to perform the sampling task nevertheless,
and produce a sample of bitstrings ${\cal S}_{\bf x}=\{\widetilde  {\bf x}^{(j)}\}$ according to these unknown
probabilities (mixed with uniform probabilities); see \cite{Google}.
It is important to realize that verifying that the quantum computer indeed performed its task of sampling from the right distribution requires to compute this distribution by a classical computer, which it cannot do for $n=53$. Therefore, the proof of quantum supremacy for large $n$ requires various extrapolation arguments
based on smaller values of $n$ or simplified circuits, in addition to  statistical reasoning.

In \cite{Google} it is assumed that the fidelity   $\phi$ is known approximately from an independent source, and part of the supremacy proof consists of
showing that the sample is indeed generated as described, which requires to estimate $\phi$.
We shall assume that for each sampled $\widetilde {\bf x}^{(j)}$ the associated probability $\widetilde {w}_j$ is known and so the sample ${\cal S}_w=\{{\widetilde {w}_j}\}$ is known. This is required for the computation of any estimator of $\phi$, both Google's and ours, apart from the estimator $T$ of Section \ref{sec:TomerT} (which requires large samples).
We emphasize that computing the probabilities  requires a classical  computation and is possible only when $n<40$ or so.

 We now generalize Google's model \eqref{eq:googsbsample} to allow for more elaborate error models discussed later.
Consider a realization of $M$ iid vectors with nonnegative components ${\bf z}_i = (z_{1i},\ldots,z_{pi})$, $i=1,\ldots, M$, distributed according to some $p$-dimensional distribution $D$ with marginal distributions $D_k$. For example, the vectors' components $z_{ki}$ can be independent, and $z_{ki} \sim D_k$, $k=1,\ldots,p$.   
For $i=1,\ldots, M=2^n$, set  
\begin{equation}\label{eq:Dirich}
{\bf w}_i = (w_{1i},\ldots,w_{pi}),\, \text{where}\,\, w_{ki}={z_{ki}}/{\sum_{j=1}^M z_{kj}},\, \text{so}\,\, \sum_{i=1}^M w_{ki}=1,  k=1,\ldots,p,
\end{equation}
and thus for each $k$, the vector $(w_{k1}, \ldots, w_{kM})$ is a \textit{random probability} vector, which has the Dirichlet(\textbf{1}) distribution  if $z_{ki} \sim$ Exp(1) iid for $i=1,\ldots,M$.  Each vector ${\bf w}_i$  is associated with a vector ${\bf x}_{{\bf w}_i}\equiv {\bf x}^{(i)} \in \{0,1\}^n$. The sampling task described next can be expressed in terms of  both ${\bf w}_i$ and ${\bf x}_{{\bf w}_i}$.

A random sample of size $N$, ${\bf x}_{{\bf \widetilde w}_1}, \ldots,{\bf x}_{{\bf \widetilde w}_N}$, or equivalently ${\bf \widetilde w}_1, \ldots,{\bf \widetilde w}_N$,
is drawn (with replacement) from the above set of  $M$ vectors ${\bf x}_{{\bf  w}_i}$, or equivalently from ${\bf w}_i$,
where draws are independent, and in each draw
the probability $\pi({\bf x}_{{\bf  w}_i}) \equiv\pi({\bf w}_i)$ of drawing ${\bf x}_{{\bf  w}_i}$ (or ${\bf w}_i$) is 
\begin{equation}
\label{eq:sbsampleZ}
\pi({\bf x}_{{\bf  w}_i}) \equiv \pi({\bf w}_i)=\phi_1 w_{1i} +\ldots +\phi_p w_{pi},\,\,\,i=1,\ldots,M,
\end{equation}
where $\phi_1,\ldots,\phi_p$ are nonnegative parameters satisfying $\phi_1+\ldots+\phi_p=1$, and so $\sum_{i=1}^M \pi({\bf w}_i)=1$. Google's model  \eqref{eq:googsbsample} with the above assumptions corresponds to \eqref{eq:sbsampleZ} with $p=2$,   $z_{ki}$$\sim$Exp(1) iid variables for $k=1$, and  for $k=2$, $z_{ki}$ is identically equal to 1.
Again, note that  the sampled $\widetilde {\bf w}_j$'s do not have the
same distribution as the original ${\bf w}_j$'s due to the sampling scheme that assigns higher probabilities to vectors with  larger components. We
denote the sample by 
\begin{equation}
\label{eq:sample}
{\cal S}_w=\{\widetilde{\bf w}_j\}=\{(\widetilde w_{1j},\ldots,\widetilde w_{pj})\},\,\,j=1,\ldots,N, 
\end{equation}
and we use ${\cal S}_{\bf x}=\{{\bf x}_{\widetilde{\bf w}_j}\}_1^N=\{\widetilde  {\bf x}^{(j)}\}_1^N$, to denote the sample, which is a multiset.
Such notation is required because the random sampling is from a finite population that is itself random,  and  $w_{kj}$ and $\widetilde w_{kj}$ are both random,  with different distributions. 
This sampling scheme is a mixture of size-biased 
sampling where in each of $N$ rounds, a 
coordinate of ${\bf w}_i$ is chosen, where $\phi_k$ is the probability of the $k$th coordinate,
and then ${\bf w}_i$ is chosen with a
probability proportional to the size of the chosen coordinate.
Section \ref{sec:BSZ} provides for more details on size bias.

Our model (1.5) is relevant for detailed modeling of the Google experiment. For a general noise model we consider $p$ possible events describing a certain malfunction in the experiment leading to a noise distribution $w_{ki} \sim D_k$ with probability $\phi_k$. In the Google experiment based on random quantum circuits it is reasonable to assume that these $w_{ki}$'s are approximately statistically independent.

In the general setup of \eqref{eq:sbsampleZ}--\eqref{eq:sample}, we study the following statistical problem: having observed the sampled
vectors $\widetilde {\bf w}_1,\ldots,\widetilde{\bf w}_N$,
one goal is to estimate $\phi_1,\ldots,\phi_p$, which
together with ${\bf w}_i$ determine the sampling scheme \eqref{eq:sbsampleZ}. Furthermore, we want to test the hypothesis that the sample is indeed generated according to the model \eqref{eq:sbsampleZ}, and also that the ${\bf w}_i$'s are generated according to the given model, which is Dirichlet distribution in Google's model \eqref{eq:googsbsample}.

     \subsection{How can such experiments be confirmed or refuted?}
     A scientific (and technological) claim such as quantum supremacy, which has
     significant implications to understanding the nature of functions that can be calculated by an effective method 
      (see \cite{NC00}, Sections 1.1.1 and 3.2.2) brings up the question of how can the evidence in the experiment be evaluated, confirmed or refuted. Since Google's quantum computer Sycamore is unique (and rather expensive), it is difficult to test it independently. This will probably hold for any similar experiments.
     
     We propose the following {protocol} for evaluation of Google's experiment; we believe that understanding  its rationale may elucidate the complexities involved in proving quantum supremacy. Independent scientists will prepare several programs (circuits) for Sycamore to be run with about $n=40$ qubits. Computing the sampling probabilities $\{w_i\}$ of these circuits  should be a task that takes several months  (on a classical computer). 
     These programs will be sent to Google for implementation.  As the implementation may somewhat change the programs due to calibration (see Section \ref{s:g1}), Google will send back the implemented programs, and large samples that they produce in a short time, which is assumed to preclude computation of the relevant $\{w_i\}$ of the implemented programs. Using classical computers the scientists will take their time and compute the set $\{w_i\}$ for each implemented programs. They will then evaluate the relation between those $\{w_i\}$'s and the samples they received. Such a protocol is likely to be relevant to other quantum supremacy demonstrations which are being pursued very actively these days.

Of course, replications of the experiment for random quantum circuits  
in the regime of 10--60 qubits, via Sycamore and and via other quantum computers, and larger sample sizes than given so far in \cite{Google} are 
necessary and valuable even if they do not follow the strict protocol above, and will give a better opportunity to examine the noise modeling.

\subsection{Paper outline}

Some background on quantum computing  is given in Section \ref{QC}.
In Section \ref{sec:BSZ} we discuss size-biased distributions and random discrete distributions, and the implications of testing goodness of fit to the size-biased sampling distribution.
In Section \ref{s:main} we concentrate on the case of $p=2$ and compare various statistical methods for estimating the fidelity.
More precisely, we analyze Google's estimator $U,$ as well as two unbiased estimators: $V$, which is a variant of $U$ that is unbiased for any given realization $\{w_i\}_1^M$,  and the
maximum likelihood estimator, denoted by MLE. We study robustness properties of estimator $U$ and related estimators, and consider a new estimator, $T$, of a different nature. The results are demonstrated briefly by simulated data, and by Google's data.
In Section \ref {s:genp} we briefly  consider estimation in the case of general $p$.
In Section \ref{sec:diffmod} we propose  more detailed noise models for
the Google sample based on the analysis of {readout} errors, and
analyze statistical estimators based on  our readout noise models.
Confidence intervals for estimated parameters are discussed in Section \ref{sec:ci}.
In Section \ref {GS} we briefly address the question of testing goodness of fit of various empirical distributions to the theoretical ones.
A preliminary study of Google's data on small circuits,  supporting our findings on the fidelity
 estimators and the relevance of our readout
 noise models,  is presented in Sections \ref{s:main} and  \ref{sec:diffmod}. Yet, neither Google's basic
   noise model nor our refined readout error model fit the observed data (Section \ref {GS}).
Section \ref{sec:conc} concludes the paper, apart from two proofs given in the Appendix.

\section {Quantum computers and Google's quantum supremacy experiment}\label{QC}

\subsection {Quantum computers}

 In this section we provide some background on quantum computing,  sampling algorithms, and Google's experiment. The reader is referred to \cite{NC00} and \cite{WS} for further information. The latter reference provides more math and physics details, and an updated summary of quantum information and computation, including various potential statistical applications.

Quantum computers are physical devices that are believed to  have the potential for solving certain computational
tasks that are well beyond the ability of classical computers. 
Shor's famous algorithm shows that if a suitable quantum computer could operate with a sufficiently low level of noise, it could
factor $n$-digit integers efficiently in roughly $n^2$ computational steps! The best-known classical
algorithms require an exponential number of steps in $n^{1/3}$. This ability for efficient factoring would allow quantum computers to break
the majority of current cryptosystems.

A {\it sampling task} is one where the computer (either quantum or classical) produces samples
from a certain probability distribution $\pi$. 
In the main example of Google's experiment paper \cite{Google}
each sample is a 0-1 vector of length $53$, where $\pi$ is a probability distribution on such vectors.
It has been hypothesized that quantum algorithms allow sampling from probability distributions well beyond the capabilities of classical
computers.

Quantum systems are inherently noisy; we cannot accurately
control them, and any interaction with them introduces further noise.  
A noisy quantum computer has the property that at every computational step 
(applying a gate, measuring a qubit) the computer makes an error 
with a certain small probability.
{\it Noisy intermediate-scale quantum (NISQ)} computers are quantum computers
where the number of qubits is in the tens or
at most in the hundreds.
Over the past decade researchers conjectured 
that the huge
computational advantage of sampling with quantum computers 
can be realized by NISQ computers that only approximate the 
target probability distribution, and predicted that this could lead to a  demonstration of quantum computational supremacy.
NISQ computers are also 
crucial to the task of creating good-quality quantum error-correction,
which is a necessary ingredient for larger-scale quantum computers.

An important feature of NISQ systems -- especially for the tasks of quantum supremacy
%and quantum error-correction
-- is the fact that a single error
in the computation sequence has a devastating effect on the outcome. In the NISQ regime,
the engineering challenge is to
keep the computation error-free. The probability
that not even a single error occurs is defined as the {\it fidelity}. 
Many companies and research groups worldwide are investing in attempts to implement quantum computations via
NISQ computers (as well as by other means).
%NISQ computers are crucial for building quantum error-correcting codes that are required for large scale quantum computers,
Statistical tools from \cite {Google} (and this paper) are relevant to the study of
fidelity and models for noise of NISQ circuits that go beyond Google's experiment. 
 
\subsection {A little more on quantum computers}

We give a brief overview of quantum computers.
A qubit is a piece of quantum memory. 
The state of a qubit can be described by a unit
vector in a two-dimensional complex Hilbert space $\cal H$. 
For example, a basis for $\cal H$ %to this Hilbert space 
can correspond to two energy levels of the hydrogen 
atom, or to horizontal and vertical polarizations of a photon. 
Quantum mechanics allows the qubit to be in a {\it superposition} of the basis vectors, described by 
an arbitrary unit vector in ${\cal H}$, and the squares of the real and complex parts of this vector represent the probabilities of the qubit to be read as 0 or 1.
The memory of a
quantum computer (quantum circuit), denoted by $C$, consists of $n$ qubits. Let ${\cal H}_k$ be the two-dimensional Hilbert space
associated with the
$k$th qubit.
The state of the entire memory of $n$ qubits is described by
a unit vector in the tensor product ${\cal H}_1 \otimes {\cal H}_2 \otimes \cdots \otimes {\cal H}_n$.
We can put one
or two
qubits through {gates}, acting on
the corresponding two- or four-dimensional Hilbert spaces, and as for classical computers,  a
small list of gates is sufficient for  universal quantum computing. 
Applying a gate amounts to applying a unitary operator on the two- or four-dimensional
space that corresponds to the qubits involved in the gate, tensored with the identity transformation 
on all other qubits. (Applying a gate thus represents a unitary transformation 
on the large $2^n$-dimensional Hilbert space that represents all the qubits.)
The system performs a given number of cycles, where each \textit{cycle} in the Google experiment consists of applying in parallel single-qubit
gates on all qubits followed by two-qubit gates on non-overlapping pairs of qubits.
The final state of the quantum computer is a unit vector
${\bf u}=(u_1,...,u_M)$ in $\mathbb C^M$, where the $M=2^n$ indexes
correspond to all 0-1 vectors of length $n$.
As already mentioned, at the end of the computation process, the state of the entire computer is  measured, giving a sample from
the probability 
distribution ${\bf \cal P}_C$ on 0-1 vectors of length $n$.

A few words on the connection between the mathematical model of quantum circuits and quantum physics:  
in quantum physics, states and 
their evolutions (the way they change in time) 
are governed by the Schr\"odinger equation. 
A solution of the 
Schr\"odinger equation can be described 
as a unitary process on a Hilbert space and 
quantum computing processes like the ones we just described form a large class of such quantum evolutions.
An interesting question is what precisely gives quantum computers their superior computing power.
The state of the quantum computer can be a huge superposition that requires an exponential number of ``amplitudes,'' 	
and therefore, quantum computers allow for specific algorithmic tasks with massive parallelism.
Another related fact is that quantum probabilities
(unlike classical probabilities) can be both positive and negative, allowing massive cancellations.

\subsection {The Google supremacy claim}
\label {s:g1}

The Google experiment is based on the building of a quantum computer (circuit) $C$ with $n$ qubits
that perform $m$ cycles of computations,  that is, rounds of  parallel operation of  a set of gates.     
At the end of the computation the qubits are  measured, leading to a sample from a probability distribution ${\cal P}_C$ on 0-1 vectors $\bf x$ of length $n$. 
This process is repeated $N$ times (for the same circuit $C$) to produce a sample of size $N$.
For the ultimate experiment ($n=53$,  with 1,113 1-qubit gates, and 430 2-qubit gates, and $m=20$ cycles) 
the quantum computer produced a sample of  several million  0-1 vectors of length 53.

The specific circuit $C$  is itself a random circuit.
For every experiment, 
specific gates are chosen at random (by a classical computer) and are fixed, thus determining (programming) the circuit. In  parts of Google's experiments (which produced most of the data), an additional \textit{calibration} was necessary after fixing the circuit, which resulted in modifying some gates and the associated circuit's probabilities.
In ideal situations, namely, without noise,
the quantum computer would produce samples from a certain probability distribution 
${\cal P}_C$
that depends on the specific circuit $C$.  When the circuit $C$ is chosen at random, the probability distribution ${\cal P}_C$
looks like an instance of the random distribution (a Porter--Thomas distribution)
 described in the introduction.

Google's quantum computer 
is ``noisy''; the size $N$ sample it produces  is modeled as follows:
a fraction $\phi$ of the samples are from ${\cal P}_C$
and a fraction $(1-\phi )$ of the samples are from a uniform distribution (or uniform noise), and $\phi$
is referred to as the {\it fidelity}.
This corresponds to \eqref{eq:googsbsample}.
We shall generalize this model in the sequel.
Google's paper made two crucial claims regarding the ultimate 53-qubit samples.

\begin {itemize}
\item[A)] The fidelity $\phi$ of their sample is above $1/1000$.

\item [B)] Producing a sample with a similar fidelity 
%statistical property 
would require 10,000 years on a supercomputer.
\end {itemize}
As it was only possible to give indirect evidence for both these claims, we shall now describe the
logic of Google's quantum supremacy argument.

For claim A) regarding the value of $\phi$, the paper describes a statistical estimator of $\phi$ 
and the argument  relies on a bold extrapolation argument that has two ingredients. 
One ingredient is a few hundred experiments in the classically tractable regime: the 
regime where the  probability distribution ${\cal P}_C$ can be computed by a classical computer and the performance of the 
quantum computer  can be tested directly. The other ingredient is a theoretical formula for computing the fidelity.
According to the Supplement to Google's paper \cite {Google}, the fidelity of entire circuits closely agrees with the prediction of the simple
mathematical formula (Formula (77) in the Supplement to \cite {Google}, 
Equation (\ref{e:77}) below).  
There are around 200 reported experiments in the classically tractable regime.
These experiments support the claim that the
prediction given by Formula (77) for the fidelity is indeed very robust and therefore may apply also
to the 53-qubit circuit in the supremacy regime.
To test whether the quantum computer produced a sample with the hoped-for properties we need to be able
to simulate the quantum computer on a classical supercomputer.
This is beyond reach for 53 qubits and therefore the samples for the 53-qubit experiments demonstrating ``supremacy'' 
are archived, but it is not possible to test them 
in any direct way. 
  
For claim B) regarding the difficulty for classical computers to compute the sampling distribution of a given quantum circuit, the Google team relies on extrapolation from the running time of a
specific algorithm they use and on results and conjectures from computational complexity 
that support the assertion that the sampling task at hand is hard, requiring an exponential number (in $n$) of computational steps on  classical computers; see Section XI in the Supplement to \cite{Google} and  \cite{AaGu19}.

A team from IBM
\cite {PGN+19} found a classical algorithm that would require only several days
of computation on a classical supercomputer (which is less than 10,000 years and
more than the running time of the quantum computer of 200 seconds). 
IBM's  and Google's algorithms computes the entire $M$ probabilities ${\cal P}_C({\bf x})$
described by the circuit.  Further
classical algorithms for performing tasks similar to Google's that might be in
conflict with the supremacy claims are given in \cite{Hua20} and  \cite {PanZha21}.
Specifically, the recent paper \cite {PanZha21}
announces a classical method for the task of producing a sample with fidelity $>$ 1/1000 for $n$=53.

\subsection {On fidelity}
\label {s:77}

The Google argument relies crucially on the following simple formula (Equation (77) in the  Supplement to \cite {Google}) 
for estimating the fidelity $\phi$ of their experiments:

\begin {equation}
\label {e:77}
\phi~\approx~ \prod_{g \in {\cal G}_1} (1-e_g) \prod_{g \in {\cal G}_2} (1-e_g) \prod_{q \in {\cal Q}} (1-e_q).
\end {equation}
Here ${\cal G}_1$ is the set of 1-gates (gates operating on a single qubit), ${\cal G}_2$ is 
the set of 2-gates (gates operating on two qubits), and
${\cal Q}$ is the set of qubits;  the term $e_g$ refers to the probability of error
of the individual gate $g$, 
and  $e_q$ is the probability of a readout error when we measure the qubit $q$.

An important aspect of \eqref{e:77} is that it is assumed that the fidelity $\phi$ is common to circuits with common  numbers of qubits and gates (and the same individual gate and qubit error rates) even if the realizations ${\cal P}_C$ vary. This allows estimation of the common $\phi$ by averaging over different  circuits with the same parameters. 

The rationale for Equation (\ref {e:77}) 
is simple: as long as there are no errors in the performance of all 
the gates and all the measurements of the qubits, the circuit produces a sample from the correct distribution. A single 
error in one of these components leads to an irrelevant sample.
The Google paper reports that for a large number of experiments the actual fidelity 
estimated by Equation (\ref {e:77})  
agrees with the statistical estimator of the fidelity up to 10\%--20\%.

A simpler form of \eqref{e:77} is obtained by replacing the detailed individual values of the fidelities
by their average value, leading to 
\begin {equation}
\label {e:77s}
\phi '\approx (1-0.0016)^{|{\cal G}_1|} (1-0.0062)^{|{\cal G}_2|} (1-0.038)^n.
\end {equation}
The Google team reports that  
$\phi'$ (more precisely, a slight variant, based on a combined estimate for 2-gate and 1-gate errors)
differs from $\phi$  by a few percent (in most cases).

We remark that the excellent predictive power of Equation \eqref {e:77} is, on its own,
a major scientific discovery as well as engineering achievement of the Google experiment. 
Concerns regarding the claimed predictive power
of Equation \eqref{e:77} were raised by
Kalai \cite {Kal}.\footnote{The same paper, like several earlier papers,
	explains Kalai's skeptical views regarding the entire endeavor of quantum computers.
	Those views, however, are not related to the present work.}  In particular, Formula \eqref{e:77} is based on the assumption of
the independence of the errors, which is often considered unrealistic in the field of system reliability
theory (see, e.g., \cite{RaHo} for a general discussion, and \cite{Devitt}  for a convenient
specific discussion of potential causes for  the dependence of errors in quantum systems).

\subsection {Google's 
	Porter--Thomas probability distributions and fidelity estimation}

In the introduction we described the Porter--Thomas distributions ${\cal P}_C({\bf x})$, 
a random probability distribution 
on a finite space $\cal X$, where  $C$ is a random circuit.
In fact,  a random circuit does not have enough randomness to generate a fully random vector
on the $(M-1)$-dimensional real simplex (which is the Dirichlet(\textbf{1})  distribution).
The vectors ${\bf \cal P}_C$
are concentrated on a tiny subset of the simplex and are thus pseudo-random. 
However, ${\bf \cal P}_C$ behaves statistically like a realization of a Porter--Thomas
distribution, and for our purposes we can consider it as random.

Once the quantum computer produces $N$ samples ${\widetilde {\bf x}}^{(1)},\dots, {\widetilde {\bf x}}^{(N)}$, the Google
estimator of $\phi$, 
\begin {equation}
\label {google-estimator}
U= 2^n\frac {1}{N}\sum_{j=1}^N {\cal P}_C(\widetilde{\bf x}^{(j)})~-1, 
\end {equation}
proposed in \cite{Google},  can be computed provided that the probabilities ${\cal P}_C(\widetilde{\bf x}^{(j)})$ are known.
This estimator is referred to in \cite {Google} as ``XEB,''  an abbreviation for
``cross-entropy benchmarking.''
A crucial aspect of this estimator and of Google's
statistical approach as a whole is that relatively small samples (of size $N$ of order $10^6$)
give powerful confirmation of the fidelity being significantly non-zero  for samples from  a huge probability space (of size $M=2^{53} \approx 10^{16}$ in Google's experiment).

We reiterate that computing the Google
estimator (\ref{google-estimator}) requires the
computation of the probabilities ${\cal P}_C(\widetilde{\bf x}^{(j)}), j=1, \dots, N$. Sampling according to the probability distribution ${\cal P}_C$ is easy for the
quantum circuit $C$, but not computing or estimating individual probabilities.
Computing ${\cal P}_C({\bf x})$
is  assumed to be
infeasible for
the ultimate experiments involving $53$ qubits, and requires heavy computations on (classical)
supercomputers for $n$ approaching $40$. It is known (see \cite{Google} and  \cite{AaGu19}) that computing ${\cal P}_C({\bf x})$, even for a 
single value $\bf x$, is NP-hard.

\section{Size bias, random  distributions, and a choice model example}	\label{sec:BSZ}
\subsection{Size-biased sampling from a random distribution}
Some background on \textit{size-biased} distributions will be useful in explaining the nature of
the sampling schemes described above. For a recent broad survey with applications see \cite{AGK}.
Let $p(x)$ be a probability function on a finite space ${\cal X} \subseteq \mathbb{R}_+ := [0, \infty]$ or a
continuous density on $\cal X$, and let $x$ be a nonnegative  random variable distributed according to $p(x)$.
We say that the random variable $x^*$ has the $p(x)$-size-biased
distribution if $P(x^*=x)= xp(x)/a$ in the discrete case, and $x^*$ has density $xp(x)/a$ in the continuous case,
where $a=Ex$ is the normalizing constant.   
If $p(x)=e^{-x}$ for $x>0$, the density of the Exp(1) distribution,
then the $p$-size-biased density is  $xe^{-x}$, corresponding to the Gamma$(2,1)$ distribution.

Consider a distribution $D$ on $\mathbb{R}_+$ with expectation =1, and a vector $(z_1, \dots ,z_M)$ whose components
are iid-distributed by $D$, and so  $Ez_i=1$.
We normalize the vector  by
setting $w_i=z_i/\sum_{j=1}^M z_j$,  and denote the normalized vector by 
${\bf w}=(w_1, \dots ,w_M)$. Since $\sum_{j=1}^M z_j/M \to 1$ with probability 1, we have $w_i \approx z_i/M$ for large $M$. The vector ${\bf w}$ defines a probability 
distribution 
on a finite set of size $M$. When $D$  is the Gamma distribution, the vector ${\bf w}$ has the Dirichlet distribution;
see, e.g., \cite {Kingman}. The special case
where the  $D$ = Exp(1)=Gamma$(1,1)$ with density $e^{-x}$ for $x>0$
arose in quantum physics; see Porter and Thomas \cite {PT}.
This case plays an important role in the Google experiment.

More generally, one can assume that ${\bf w}$ 
is a realization
of some distribution on the $(M-1)$-simplex, that is, a random vector of nonnegative components summing to one; this structure is known as a random discrete distribution; see Kingman \cite {Kingman}.

\begin{proposition}
	\label{prop{sizb}}
	Let $z_1,z_2,...$ be nonnegative, iid-distributed by $D$, with expectation =1, and for each $M$ define $w_i^{(M)}=z_i/\sum_{j=1}^M z_j$. Let $x^*_M$ denote a value drawn  at
	random from   $\{z_1,\ldots,z_M\}$, where $P(x^*_M=z_i) = w_i^{(M)}$, $i=1,\ldots,M$. Then with
	probability =1 (over sequences $z_1,z_2,...$) we have that $x^*_M$ converges in
	distribution to the $D$-size-biased distribution as $M \to \infty$.  
\end{proposition}
\textit{Proof}:
Let $F_n$ be the empirical distribution function   of $z_1,\ldots,z_M$ (which assigns probability $1/M$ to each $z_i$). If  $x^*$ takes one of the values $z_1,\ldots,z_M$ with probability  $P(x^*=z_i) ={z_i}/{\sum_{j=1}^M z_j}$, then clearly the distribution of $x^*$ is the $F_n$-size-biased distribution. 
Since $F_n$ converges in distribution to $D$, the $F_n$-size-biased distribution converges in distribution
to the $D$-size-biased distribution by Theorem 2.3 in \cite{AGK}. \qed

In the sampling scheme of \eqref{eq:googsbsample}, $N$ values of $w_i$ are sampled from the
set $\{w_i\}_1^M$, where with probability $\phi$ the value $w_i$  is drawn with the
probability $w_i^{(M)}$ above. The corresponding values $z_i$ are asymptotically  distributed according
to a mixture of $D$-size-biased values with weight $\phi$, and the original $D$ with weight $1-\phi$ (since then a value of $z$ is chosen from iid $z_i \sim D$ with equal probabilities). Note that since ${\frac{1}{M}\sum_{j=1}^M z_j} \approx 1$ by the law of large numbers, $w_i^{(M)} \approx z_i/M$.  Obviously,
relative to the distribution $D$, large values of $z_i$ or $w_i$ are overrepresented under size-biased sampling, and
the distribution of the sample is tilted to the right.

%
%When $D$ is Exp(1), that is, the original $z$'s have the Exp(1) distribution, the distribution of the sampled $z_i$ is a mixture of approximately Gamma(2,1) (with density function $ze^{-z}$)  with probability $\phi$  and the original $D$ (with density  $e^{-z}$) with probability $1-\phi$. 

\subsection{ Testing the size-biased distribution and model \eqref{eq:googsbsample}}\label{sec:wvsx}
The quantum computer produces a sample of bitstrings ${\cal S}_{\bf x}=\{\widetilde {\bf x}_j\}_1^N$.
The $\{w_i\}$'s associated with the  bitstrings ${\bf x}_i$ must be computed by a classical computer, and  with the known association one can focus on the sample ${\cal S}_w=\{\widetilde {w}_j\}$, and 
${\cal S}_{\bf x}$ is now of no further use. 

We can test whether $\{\widetilde w_j\}$ arise from a mixture of Dirichlet and size-biased Dirichlet variables according to  \eqref{eq:googsbsample} (call the distribution of this mixture $\cal D$), by, say, looking at a histogram of the sample ${\cal S}_w$ and comparing it to the theoretical mixture density of $\cal D$, which we do in Figure \ref{fig:hists} of Section \ref{GS}. If the model fits, it could be because  Google's sampling model \eqref{eq:googsbsample}
indeed  holds with the given $\{w_i\}$'s.  In this case the quantum computer produces a sample which is correlated with the $w_i$'s, that is,  the``signal" can be identified from the noisy  sample, indicating {supremacy} if $n$ is large. If  \eqref{eq:googsbsample} is applied with a different (independent) set of probabilities $\{w'_i\}$,
the histogram of the sample ${\cal S}_w$ will not exhibit size bias.  Now, consider a statistic like $U=\frac{1}{N} M\sum_{i=1}^N \widetilde w_i-1$ of  \eqref{google-estimator-i}. 
Under the model (1.1)  
we expect to observe large  $\widetilde w_i$'s  due to the size bias, and therefore $EU>0$; thus
a positive significant $U$ demonstrates the presence of the the $w_i$'s, i.e., supremacy (if $n$ is large).

However, if the  sample ${\cal S}_w$  exhibits size bias, it does not prove that the model \eqref{eq:googsbsample} holds as stated. For example,
if some bitstrings are rejected from the sample with probabilities  that are independent of the $w_i$'s, and hence some of the sampled $\widetilde w_j$'s
are rejected, then the fit to $\cal D$ will still hold. This will be explained in Section \ref{sec:Rob}. 
Moreover, a fit to  $\cal D$ will be exhibited also if \eqref{eq:googsbsample}  is replaced by  $\pi({\bf x}^{(i)}) = \pi({ w}_i)=\phi {w_{i}}+(1-\phi)v_i$
with a set of probabilities $\{v_i\}$ that are not uniform, provided that they are independent of the $w_i$'s, that is, the nature of the noise is unimportant
as long as the noise is independent of the probabilities $w_i$. 
Robustness of $U$ under the  above rejection model and   to different kinds of noise instead of uniform noise are discussed in Section \ref{sec:Rob}.

Thus the  fit of histograms to $\cal D$ and a positive significant $U$ 
can indicate supremacy, without confirming the model \eqref{eq:googsbsample}.
To test the model \eqref{eq:googsbsample} we need much data; we can then  compute a chi-square goodness of fit statistic, for example; see Section \ref{GS}.

Contrary to Google's $U$, the estimator $T$ of $\phi$, considered in Section \ref{sec:TomerT},   is nearly unbiased if model \eqref{eq:googsbsample} holds, but is not robust to sampling with rejection, for example. If one knows or claims a certain value of $\phi$, and $T$ suggests a different value, it could be because of certain deviations from \eqref{eq:googsbsample}, which may not affect $U$. Computing the estimator $T$ does not require knowledge of $w_i$'s (unlike $U$ and chi-square), but does require large samples relative to $M$.

\subsection{An individual choice model}

We briefly describe 
a simple application, unrelated to quantum computing, that is a variation on problems in individual choice
models in economics; see, e.g., \cite{Hensher} for a simple introduction. It provides another 
perspective on the models we discuss, and suggests other potential applications. 
Consider $M$ items, say  refrigerators (fridges), in a store  (they need not be all different).
Each fridge is characterized by $p$ attributes such as size,  price per given size, energy consumption per size,  etc.,  which for fridge $i$ are $w_{1i}, \ldots, w_{pi}$.  By considering one attribute relative to another (a further example could be price per size and energy consumption) we attempt to make the attributes $(w_{1i}, \ldots, w_{pi})$ independent. Each of $N$ customers chooses one fridge and it is assumed that the probability of choosing fridge $i$ is given by  \eqref{eq:sbsampleZ}, that is, a convex combination of the $w$'s that characterize this fridge with the $\phi$'s as coefficients.  Thus,  the vector of $\phi$'s characterizes the population of customers from which we have a sample of $N$. The parameter $\phi_k$ quantifies the  weight the population assigns to attribute $k$ when choosing a fridge, and this is a quantity the seller  or the producer would like to know. This paper proposes various ways of estimating the $\phi$'s from the sample. The information consists of the $N$ chosen fridges and their attributes, but the customers do not reveal their selection process, and they may not be aware of it.

A useful interpretation of the linear combination in \eqref{eq:sbsampleZ} is the following: each customer chooses an attribute  at random with probabilities according to the  $\phi$'s, and then if attribute $k$ was chosen by customer $i$, she chooses fridge $i$ with a probability proportional to $w_{ki}$, and hence it is a size-biased selection. (See for  example \cite{Chan},  where
size-biased observations with respect to one  variable out of several are considered. See also Section \ref{sec:BSZ}.) Equivalently, a proportion $\phi_k$ of customers choose  in proportion to the value of the $k$th attribute.  In either interpretation, $\phi_k$ represents the importance of attribute $k$, thus quantifying the relative  importance  of attributes  like size, price, energy consumption, etc., in the item selection.

\section{Estimation of the fidelity $\phi$ for $\lowercase{p} =2$}\label{sec:estfi}
\label {s:main}

We first discuss the estimation of $\phi$ given a sample $\widetilde w_1,\ldots, \widetilde w_N$ according to Google's model \eqref{eq:googsbsample}. 
In Section \ref{s:genp} we  generalize to estimation of $\phi_1, \ldots,\phi_p$ in the case of  observing  $\widetilde {\bf w}_1, \ldots, \widetilde{\bf w}_N$, sampled according to \eqref{eq:sbsampleZ}. 
We study two types of analyses and consider biases, variances, etc.: the first type is conditioned on $w_1,\ldots,w_M$, which depend on the quantum circuit, and the second is with expectation taken over the random $w_1,\ldots,w_M$ when considering random circuits. We compare these analyses and their implications for different estimators, and for different parameters of the problem. In particular, we compare Google's estimator with the maximum likelihood estimator (MLE). 

Google's estimator $U$, defined in \eqref{google-estimator-i} and again in \eqref{eq:linmestgoog} below, and the MLE, defined in \eqref{eq:flik}, depend only on the sampled
$\widetilde {\bf w}_1, \ldots, \widetilde{\bf w}_N$. Therefore, they
are statistics (i.e., functions of observed data) if we assume that the sample ${\cal S}_w=\{{\widetilde {w}_j}\}_1^N$
is observed. The estimator $T$ given in Section \ref{sec:TomerT} requires only knowledge of the sample of bitstrings ${\cal S}_{\bf x}=\{\widetilde{\bf x}^{(j)}\}$ (and not of the associated probabilities). 
Another estimator considered below in \eqref{eq:GV}, $V$, is not a statistic in the sense that computing it requires the knowledge of $w^{(2)}:=\sum_{i=1}^M w_i^2$. We mention again that a quantum circuit produces a sample of bitstrings ${\bf x}_{\widetilde { w}_1}, \ldots, {\bf x}_{\widetilde {w}_N}$; however, calculating the associated  $\widetilde { w}_1, \ldots, \widetilde{w}_N$ is done by classical computers only, and for limited values of $M=2^n$. For $n<40$, a classical supercomputer can compute all $w_i$ and hence $V$ can also be computed. For $n>50$ or so, computing the $w_i$'s is considered practically impossible, and hence  the statistics presented in \cite{Google} and here (with the exception of $T$) cannot be computed. Therefore, Google's supremacy claim with $n=53$ is based on skillful extrapolations. 

Google's estimator $U$ turns out to be biased for each given circuit. Averaged over sufficiently many different circuits, or in expectation over the random probabilities $w_{\bf x}={\cal P}_C({\bf x})$,  it becomes unbiased, and hence consistent as a function of the number of circuits, as we show below. The bias for each circuit increases the variance between circuits; this variance diminishes along with the bias as $\phi$ becomes small for large $M$ and negligible for $M=2^{53}$.

\subsection{Moment estimators}
\label {ME}

Google's estimator (see \eqref{google-estimator-i} or \eqref{google-estimator}), is basically a moment estimator, a notion we briefly review.
The \textit{method of moments} (see, e.g., \cite{Greene}, Chapter 18) can be described as follows: let $x_i$ be $N$ iid observations taking values
in some space $\cal X$ and assume that $x_i \sim F_\theta$, where $\theta \in \Theta \subseteq \mathbb{R}^p$. Let $\varphi : {\cal X} \to \mathbb{R}^p$ and
assume that $E_\theta \varphi(x_i)=g(\theta)$, where $g : \Theta \to \mathbb{R}^p$ is one-to-one.
% so that $g^{-1}(g(\theta))=\theta$. 
A moment estimator of $\theta$ is of the form 
\begin{equation}
\label{eq:momest}
T_N= g^{-1}\left(\frac{1}{N}\sum_{i=1}^N \varphi(x_i) \right).
\end{equation}  
Under standard assumptions, the law of large numbers implies that $\frac{1}{N}\sum_{i=1}^N \varphi(x_i) \to E_\theta \varphi(x_i)$ as $N \to \infty$, and
further standard smoothness assumptions imply that  $T_N$ is \textit{consistent}; that is, $T_N$ converges to $\theta$ in probability; moreover, $\sqrt{N}(T_N-\theta)\to N({\bf 0}, \Sigma)$ in distribution,
where the asymptotic covariance matrix $\Sigma$ depends on the variance of the variables $x_i$, the functions $\varphi$ and $g$, and their derivatives.
In this case we say that $T_N$ is \textit{root-$N$ consistent} and \textit{asymptotically normal}.

In Section \ref{sec:condMLE} we shall discuss the \textit{maximum likelihood estimator} (MLE). The MLE is known to be \textit{asymptotically efficient}, that is, asymptotically normal with the
smallest asymptotic variance among asymptotically normal  estimators, under certain regularity conditions (see, e.g., Theorem 3.10, Chapter 6 in \cite{LehCas}). 
%(in terms of having the
%sma variance) harder to compute
Theorems 4.3 and 5.3 of Chapter 6 in \cite{LehCas} show that one Newton--Raphson iteration (toward the MLE)
starting  from any root-$N$ consistent  estimator yields an asymptotically efficient estimator, which
means that like the MLE it  has the
smallest asymptotic variance.

\subsection{Google's estimator $U$ of the fidelity  and its unbiased variant $V$}\label{sec:Gphi}
Let $E^w$ denote the conditional expectation of some function of $\{\widetilde w_j\}_1^N$ given $\{w_i\}_1^M$, where $\{\widetilde w_j\}_1^N$  are sampled according to \eqref{eq:googsbsample}. Expectation over the randomness of $\{w_i\}_1^M$ is denoted by $E$. Thus $Eg(\widetilde w_j)=EE^wg(\widetilde w_j)$ denotes expectation over both the sampling of $\widetilde w_j$ and the randomness of the random probabilities $w_i$.
Similarly $Var^w$ denotes variance  conditioned on $\{w_i\}_1^M$.	

In this section we discuss Google's estimator of $\phi$ for the sampling distribution of \eqref{eq:googsbsample}.
Consider a vector  $(w_1, \ldots,w_M)\sim$ Dirichlet$(\bf 1)$, which can be obtained as mentioned before by $w_i=z_{i}/\sum_{j=1}^M z_j$ with iid $z_i \sim$ Exp(1). Let $\widetilde w_{1},\ldots,\widetilde w_{N}$  be
a sample (with replacement) of $N$ values of $w_i$ that are sampled independently according to the mixture probabilities 
\begin{equation}
\label{eq:glik}
\pi({w}_i)=\phi w_{i}+(1-\phi) /M.%=%\pi({z}_i)=\phi z_{i}/M\bar z+(1-\phi) /M.
\end{equation}  Google's estimator $U$ of $\phi$  is (in our notation)
\begin{equation}\label{eq:linmestgoog}
U=\frac{1}{N}\sum_{j=1}^N
M\widetilde w_{j}-1 .
\end{equation}
We now compute its expectation and variance conditioned on  $(w_1, \ldots,w_M)$. 
Put $ w^{(k)}=\sum_{i=1}^M w_i^k$ and so  $Ew^{(k)}=k!/[(M+1)\cdots(M+k-1)]$  by \eqref{eq:mom}, and in particular $Ew^{(2)}=2/(M+1)$. 

With expectations being conditional on $w_i$ we have, using $\sum_{i=1}^M w_i=1$,
\begin{equation}\label{eq:toEU}
E^wM\widetilde w_{i}= \sum_{i=1}^M M w_{i}
[\phi w_{i}+(1-\phi)/M]= \phi(M w^{(2)}-1)+1.
\end{equation}
\textbf{Remark:} If \eqref{eq:glik} is replaced by 
%\begin{equation}
%\label{eq:glik}
$\pi({w}_i)=\phi w_{i}+(1-\phi)v_i$ for some exchangeable random $\{v_i\}$ (with $\sum_{i=1}^M v_i=1$ and therefore $Ev_i=1/M$) independent of $\{w_i\}$, then $E^wM\sum_{i=1}^M w_iv_i=M\sum_{i=1}^M w_iEv_i=1$ and we obtain the same result as \eqref{eq:toEU}, showing that $U$ is robust against changing the sampling probabilities under errors. Robustness properties of $U$ and other estimators are discussed in Section \ref{sec:Rob}.
The $v_i$'s do affect the variance of $U$; however, we shall not pursue these variance calculations.

Returning to \eqref{eq:toEU} we obtain
\begin{equation}\label{eq:EEU}
E^wU = \phi(M w^{(2)}-1).
\end{equation}
Thus we see that for a given set  $\{w_i\}_1^M$, that is, a  given circuit, Google's estimator $U$ is biased. Clearly, relative to $\phi$, the bias converges to zero  as $M \to \infty$; see below. However, the bias does not depend on the sample size $N$, and therefore as $N \to \infty$ the estimator $U$ does not converge to $\phi$; that is, it is inconsistent.

Google's estimator $U$ can be improved and rendered consistent in $N$ by 
adjusting it  to be unbiased.  Define 
\begin{equation} \label{eq:GV} V=U/(M w^{(2)}-1).
\end{equation}
An averaged version of $V$ (that is still biased and hence not consistent) appears in Equation (21) in the Supplement to \cite{Google}.
We shall see that the unbiased estimator $V$ has a smaller variance than that of $U$, and thus it is an improvement upon  $U$.  
Unlike $U$, computing the estimator $V$ requires access to all $\{w_i\}_1^M$ so that we can compute $w^{(2)}$, and hence it is not a statistic relative to the sample $\widetilde w_{1},\ldots,\widetilde w_{N}$. 

Recall that $(w_1, \ldots,w_M)$ has the Dirichlet distribution; however, the calculations that follow require only conditions on moments up to order 4.
Taking expectation over $\{w_i\}_1^M$ and using $Ew^{(2)}=2/(M+1)$,  we get
\begin{equation}\label{eq:EUUU}
EU=EE^wU = E[\phi(M w^{(2)}-1)]= \phi\left(\frac{M-1}{M+1}\right) \approx \phi,
\end{equation}
and the law of large numbers implies that $U \to \phi\left(\frac{M-1}{M+1}\right)$ almost surely as $N \to  \infty$. 
Thus, $U$ is nearly unbiased for large $M$ when considered over the randomness of both the sampling process of \eqref{eq:googsbsample} and the random probabilities $\{w_i\}_1^M$, and can be adjusted by dividing it by $(M-1)/(M+1)$ to make it unbiased and consistent as $N \to \infty$.
Thus, if $U$ is computed from  samples taken from different independent  sets of $\{w_i\}_1^M$ (that is, different independent circuits) and the results are averaged, then the resulting average is nearly unbiased for large $M$; however, as we shall see, the bias of $U$ for each realization $\{w_i\}_1^M$ makes its variance larger than that of $V$.

To obtain the variance we compute $E^wM^2\widetilde w_{i}^2=M^2\sum_{i=1}^M w_{i}^2
[\phi w_{i}+(1-\phi)/M]=\phi (M^2 w^{(3)}-M  w^{(2)})+M  w^{(2)}$. Therefore,
\begin{equation} 
\begin{split}\label{eq:varUU} Var^wU&=\frac{1}{N}Var^w(M\widetilde w_{i})\\=&\frac{1}{N}\left[\phi (M^2  w^{(3)}-M  w^{(2)})+M  w^{(2)} - (\phi(M  w^{(2)}-1)+1)^2\right]\\
=&\frac{1}{N}\left[\phi(M^2  w^{(3)}-3M  w^{(2)}+2)-\phi^2(M  w^{(2)}-1)^2+M  w^{(2)}-1\right].
\end{split}
\end{equation}
Taking the denominator  $(M w^{(2)}-1)$ of \eqref{eq:GV} into account we obtain
\begin{multline}\label{eq:varv}
Var^w(V)\\=\frac{1}{N(M w^{(2)}-1)^2}\left[\phi(M^2  w^{(3)}-3M  w^{(2)}+2)-\phi^2(M  w^{(2)}-1)^2+M  w^{(2)}-1\right].
\end{multline}
Given all $\{w_i\}_1^M$ of a given circuit, or just $w^{(2)}$ and $w^{(3)}$, the quantities of \eqref{eq:varUU} and \eqref{eq:varv} can be computed.

Under the moment conditions of \eqref{eq:mom} we have $w^{(2)} \approx 2/M$ and $w^{(3)} \approx 6/M^2$ (where  the approximation is in the sense that the ratio of the two sides converges to 1 as $M \to \infty$), and therefore 
\begin{equation}\label{eq: VarianceV}
Var^w(U)\approx	Var^w(V) \approx \frac{1}{N}(2\phi - \phi^2+1).
\end{equation}
The difference between the two estimators $U$ and $V$  is that $V$ is unbiased and $U$ is not, and the overall (unconditional) variance of $U$ depends also on $Var(E^wU)$, which we discuss next.
Equation \eqref{eq: VarianceV} provides $Var^w(U)$,  the conditional variance of $U$ of \eqref{eq:linmestgoog}.
Recalling the Pythagorean formula, aka the law of total variance, we have (for any $U$)
\begin{equation}
\label{eq:pytha} Var(U)= E[Var^w(U)]+Var[E^w(U)],
\end{equation} 
and given the fact that $V$ is conditionally unbiased, that is, $E^w(V)=\phi$ and hence $Var[E^w(V)] =0$, we conclude by \eqref{eq: VarianceV} that 
\begin{equation}\label{eq:varrvv}
Var(V)=EVar^w(V)\approx \frac{1}{N}(2\phi - \phi^2+1).
\end{equation} 

The situation with $U$ is different and in view of \eqref{eq:pytha}
we have to compute $Var(E^wU)$. By \eqref{eq:mom} $Var(w_i^2)=Ew_i^4-(Ew_i^2)^2 \approx 4!/M^4-(2/M^2)^2=20/M^4$, and
by \eqref{eq:EEU} we have
$Var(E^wU) = Var[\phi(M w^{(2)}-1)] \approx  20\phi^2/M$. From \eqref{eq:pytha} and \eqref{eq: VarianceV} we conclude that the overall variance of $U$ is 
\begin{equation}\label{eq:varruu}
Var(U)\approx \frac{1}{N}(2\phi - \phi^2+1)+ 20\phi^2/M.
  \end{equation}
This variance does not decrease to zero when the sample size $N \to \infty$, and hence $U$ as a function of $N$ is not a consistent sequence of estimators. If $M$ and $N$ are of similar order, the term $20\phi^2/M$ may matter and $U$ will be inferior to $V$ (and to the MLE as shown later). This may happen if one extrapolates from relatively small circuits, that is, small $M$. When $n=53$, the number of qubits in Google's ultimate quantum computer, and when $N$ is of order $10^6$, a typical sample size in Google's experiment, that term is  unlikely to matter.

The variances of $V$ and $U$ given in \eqref{eq:varrvv} and \eqref{eq:varruu}
are approximations to the respective variances when 
$\{w_i\}_1^M$ are considered random with the first four moments corresponding to the Dirichlet distribution. 
Suppose that we have $L$ estimators $V_\ell$ arising from different circuits having a common fidelity $\phi$. Then for $\overline V=\frac{1}{L}\sum_{\ell=1}^L\!V_\ell$ we have $E \overline V=\phi$, and by \eqref{eq:varrvv}  $Var(\overline V)\approx \frac{1}{LN}(2\phi - \phi^2+1)$. A similar result holds for $U$, where by  \eqref{eq:EUUU}  we obtain for $\overline U=\frac{1}{L}\sum_{\ell=1}^L\!U_\ell$  that $E \overline U \approx \phi$, and by \eqref{eq:varruu} that $Var(\overline U)\approx \frac{1}{LN}(2\phi - \phi^2+1)+\frac{20}{LM}\phi^2$. These results will be used in Section \ref{sec:ci} for constructing approximate confidence intervals.

\subsection{A robustness property of $U$}\label{sec:Rob}
Consider sampling according to \eqref{eq:glik}, but suppose that certain bitstrings ${\bf x}_i$ and hence the associated $w_i$ are rejected from the sample with  probability $\rho_i \in [0,1]$, and hence are accepted with probability $\tau_i:=1-\rho_i$, where the  $w_i$'s are iid as usual and are assigned to the $\tau_k$'s at random. Sampling continues until $N$ bitstrings are obtained.   We show that  Google's estimator $U$ as well as $V$ and the MLE  remain valid for estimating $\phi$ and there is no need for adjustments, nor to know the acceptance probabilities $\tau_i$. Conditioning on the $w_i$'s and attaching the $\tau_k$'s to them at random we have 
instead of \eqref{eq:glik} the sampling probabilities 
\begin{equation*}
\label{eq:gliktau}
\pi_\tau({w}_i)=\,\sum_{k=1}^M\tau_k[\phi\,w_{i}+(1-\phi) /M]\big/\big\{\sum_{i=1}^M\sum_{k=1}^M\tau_k[\phi\,w_{i}+(1-\phi) /M]\big\}=
\phi\,w_{i}+(1-\phi) /M.
\end{equation*}
Thus the sampling distribution of a bitstring is the same as that of \eqref{eq:glik} and hence the estimators of Section \ref{sec:estfi} continue to be valid. However, once sampling is repeated with the same (random) association of the $\tau_k$'s to the $w_i$'s, the sampled $\widetilde w_i$ are not independent. If only two of the $\tau_k$'s are positive, say, then  observing $\widetilde w_i$ suggests that it is associated with the bigger $\tau_k$ and is more likely to appear again. However, the dependence vanishes as $M \to \infty$ if, for example, half of the $\tau_i$'s are equal to 1 and  half to zero. In this case, for large $M$, we simply have to replace $M$ by $M/2$ in the formulas of Section \ref{sec:Gphi}. A similar result holds if for some $c, d \in (0,1]$ we have that $cM$ of the 
$\tau_i$'s satisfy $\tau_i>d$. For a discussion of the robustness of $U$ against changing the sampling probabilities under error from a uniform to a general distribution (independent of $\{w_i\})$  see the remark following \eqref{eq:toEU}. Like the acceptance-rejection model above, the nonuniform  probabilities entail dependence, which affects the variance of $U$. 

\subsection{Maximum likelihood estimation of $\phi$}\label{sec:condMLE}

The likelihood function of  the sample $\widetilde {w}_1, \ldots, \widetilde{w}_N$ is given by the product $\prod_{j=1}^N ({\phi \widetilde w_{j}+(1-\phi)/M})$, and the MLE of $\phi$ is obtained by maximizing it with respect to $\phi$. This can be done by equating the derivative of the log-likelihood to zero; that is,  
the {MLE} is the solution in $\phi$ to the equation 

\begin{equation}\label{eq:flik}
f(\phi):=\frac{\partial}{\partial \phi}\sum_{j=1}^N\log({\phi \widetilde w_{j}+(1-\phi)/M})	=\sum_{j=1}^N \frac{\widetilde w_{j}-1/M}{\phi \widetilde w_{j}+(1-\phi)/M}=0.
\end{equation}
An algorithm for the solution will be given below. 	
The MLE for several different circuits together is considered in Equation (20)  in the Supplement to \cite{Google}. 

We now discuss the variance of the MLE. Under the present conditions, it is well known that the asymptotic (in $N$)  variance of the MLE is given by $1/\mathscr{I}(\phi)$, where $\mathscr{I}(\phi)$ is 	the Fisher information, which is the expectation, conditioned on $\{w_i\}_1^M$, of the second derivative of the log-likelihood function with a negative sign. See, e.g., \cite{LehCas} Chapter 6, Theorem 3.10. We have 
\begin{multline}\label{eq:Fisher}
\mathscr{I}(\phi)=	-E^w\left[\frac{\partial}{\partial \phi}\sum_{j=1}^N \frac{\widetilde w_{j}-1/M}{\phi \widetilde w_{j} +(1-\phi)/M}\right]=
NE^w\frac{(\widetilde w_{j}-1/M)^2}{[\phi \widetilde w_{j} +(1-\phi)/M]^2}\\=N\sum_{i=1}^M \frac{(w_{i}-1/M)^2}{[\phi w_{i} +(1-\phi)/M]^2}[\phi w_{i}+(1-\phi)/M]
=N\sum_{i=1}^M \frac{( w_{i}-1/M)^2}{\phi  w_{i} +(1-\phi)/M}.
\end{multline}
Given all $\{w_i\}_1^M$ of a specific circuit and the fidelity $\phi$, the Fisher information can be computed. When $\phi$ is unknown, plug-in estimation is used. 

The  quantity in \eqref{eq:Fisher} can be approximated by $N\int_0^\infty \frac{(z-1)^2}{\phi z +1-\phi}e^{-z}dz$ if $M$ is large, and we approximate $w_i$ by $z_i/M$ and assume that  $z_i \sim$ Exp(1). Recalling that the asymptotic variance of the MLE is $1/\mathscr{I}(\phi)$, we have
\begin{equation} \label{eq:varMLE}  Var^w(MLE)\approx\frac{1}{N\int_0^\infty \frac{(z-1)^2}{\phi z +1-\phi}e^{-z}dz}.
\end{equation}
In view of \eqref{eq:pytha} we need to consider $Var[E^w(MLE)]$. In our case the variables, the log-likelihood, and its derivatives are all bounded and therefore
$E^w[(MLE)-\phi]=O(1/N)$ (see, e.g., \cite{CH}, Section 9.2) and therefore its variance is of order $O(1/N^2)$, which we neglect, and our approximation for $Var(MLE)$, the unconditional variance, is the same as in \eqref{eq:varMLE}. 

If we have  $L$ independent MLE estimators, denoted by MLE$_\ell$, from $L$ independent samples $\{\widetilde w_j\}$,  arising either from the same or from several circuits assuming that they have a common $\phi$, we can  define $\overline{MLE} =\frac{1}{L}\sum_{\ell=1}^L MLE_\ell$. We then have 
\begin{equation} \label{eq:varMLEofL}
Var(\overline{MLE})\approx\frac{1}{LN\int_0^\infty \frac{(z-1)^2}{\phi z +1-\phi}e^{-z}dz}.
\end{equation}
It is easy to see that computing the MLE from joint likelihood of all $L$ independent samples requires summing \eqref{eq:flik} over the $L$ samples, and  the Fisher information will be the sum of $L$ terms that are similar to \eqref{eq:Fisher}, and the resulting approximation of the variance of the joint MLE  coincides with \eqref{eq:varMLEofL}. 

We next provide explicit formulas for the Newton--Raphson algorithm for computing the MLE. This requires computations of 
\begin{equation}\label{eq:Hess}
J(\phi)=\frac{\partial^2}{\partial \phi^2}\sum_{j=1}^N\log({\phi \widetilde w_{j}+(1-\phi)/M})=-\sum_{j=1}^N \frac{(\widetilde w_{j}-1/M)^2}{[\phi \widetilde w_{j} +(1-\phi)/M]^2}.
\end{equation}
With $f(\phi)$ of \eqref{eq:flik} the Newton--Raphson iterations are given by 
\begin{equation}\label{eq:NewtonRaph}
\phi_{k+1}=\phi_{k}-f(\phi_{k})/J(\phi_{k}).
\end{equation}

To see the relation between the MLE and the estimator $V$, we write a first-order Taylor expansion of the function $f(\phi)$ of \eqref{eq:flik} at $\phi=0$  to obtain
$$f(\phi)=\sum_{j=1}^N \frac{\widetilde w_{j}-1/M}{\phi \widetilde w_{j}+(1-\phi)/M}\approx M\sum_{j=1}^N (\widetilde w_{j}-1/M)-\phi M^2\sum_{j=1}^N {(\widetilde w_{i}-1/M)^2}.$$ 
By \eqref{eq:glik} we have
$$M^2\frac{1}{N}\sum_{j=1}^N {(\widetilde w_{j}-1/M)^2} \approx M^2E^w{(\widetilde w_{j}-1/M)^2}=M^2\sum_{i=1}^M {(w_{i}-1/M)^2}[\phi w_{i}+(1-\phi)/M],$$
and for $\phi=0$ the latter expression equals $Mw^{(2)}-1$ since $\sum_{i=1}^M w_i =1$. Thus, setting $f(\phi)=0$ in the above Taylor expansion and solving for $\phi$, we obtain that the solution is  $(\frac{1}{N}\sum_{i=1}^N {M\widetilde w_{i}-1) }/(Mw^{(2)}-1)$, which is the estimator $V$, showing that $V$ is an approximation to the MLE for small values of $\phi$.

\subsection{Estimator comparisons}
\begin{figure}[h]
	\begin{center}
		\includegraphics[width=13.2cm, height=9.1cm]{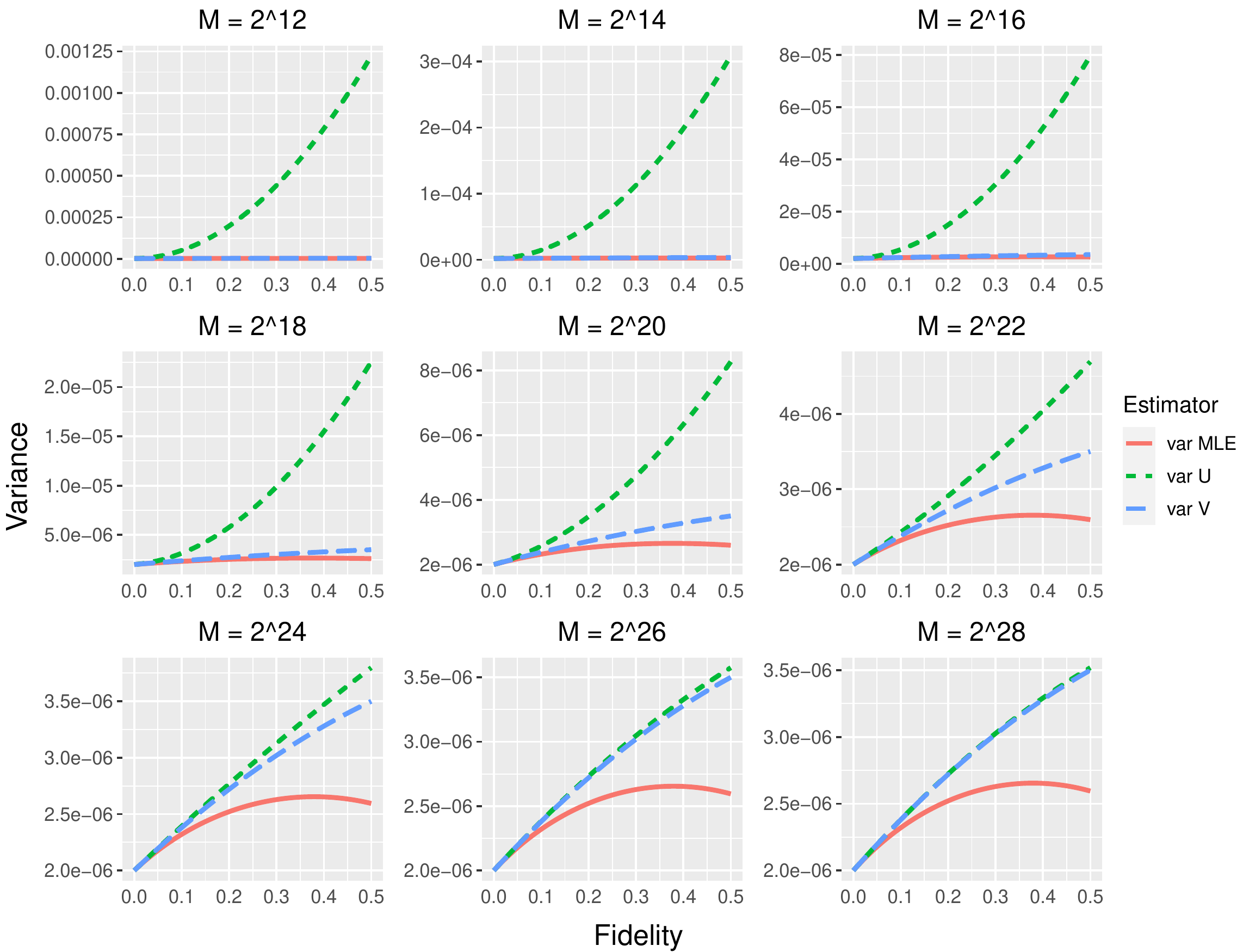}
		\caption{Variances of MLE, $U$, and $V$ as a function of $\phi$.}
		\label{fig:curvUVM}
	\end{center}
\end{figure}

We  compare the unconditional variances of the estimators $U$, $V$, and MLE for $N = 500,\!000$,\, $M = 2^{12} - 2^{28}$,  and $0 \le \phi \le 1/2$ according to the formulas 
\begin{align*} Var(U) \approx& \frac{1}{N}(2\phi-\phi^2+1) + \frac{20\phi^2}{M}, \quad Var(V) \approx \frac{1}{N}(2\phi-\phi^2+1), 
\\
&Var(MLE)\approx \frac{1}{N\int_0^\infty \frac{ (z-1)^2}{\phi z + 1 - \phi}e^{-z}dz}.
\end{align*}
A second-order Taylor expansion of the latter approximation of $Var(MLE)$ around $\phi=0$ yields $\frac{1}{N}(2\phi-5\phi^2+1)$, which is smaller than the approximate expression of $Var(V)$; however, we see again that for small $\phi$ these variances are close together.
The variances are given as a function of $\phi$ in Figure \ref{fig:curvUVM}. Note that the ranges of the  y-axes  change between plots in Figure \ref{fig:curvUVM} and some of the other figures. Consider the following numerical example: for $M=2^{18}$ and $\phi=0.15$ we have that $Var(V)$ based on about $N=299,\!000$ equals $Var(U)$ based on $N=500,\!000$, and so $V$ entails a reduction by 40\% in the required sample size to achieve the same variance or confidence interval width.

\begin{figure}[h]
	\begin{center}
		\includegraphics[width=13.5cm, height=3.5cm]{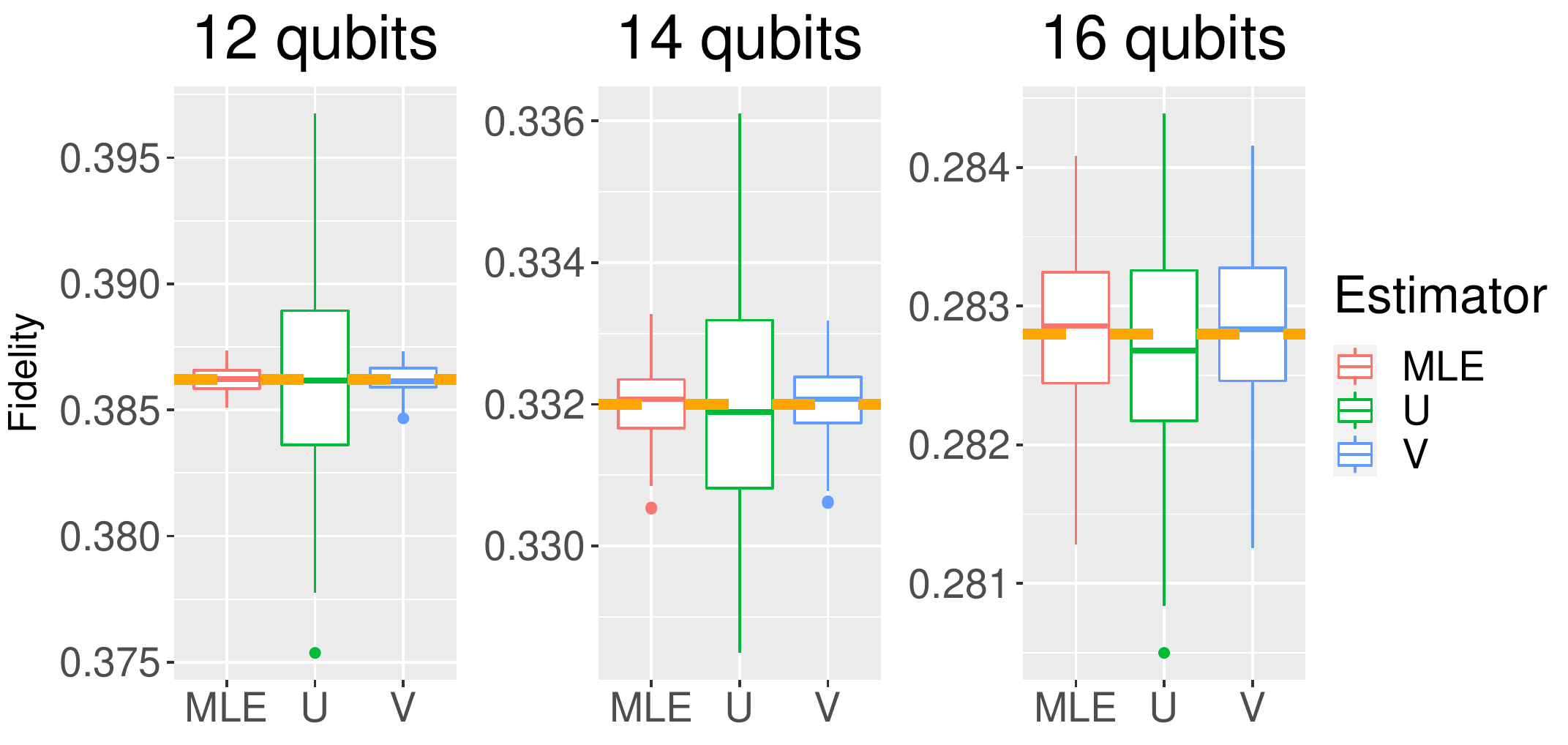}
		\caption{Boxplots of 100 averages of MLE, $U$, and $V$.}
		\label{fig:box}
	\end{center}
\end{figure}

\begin{figure}[h]
	\begin{center}
		\includegraphics[width=13.2cm, height=6.2cm]{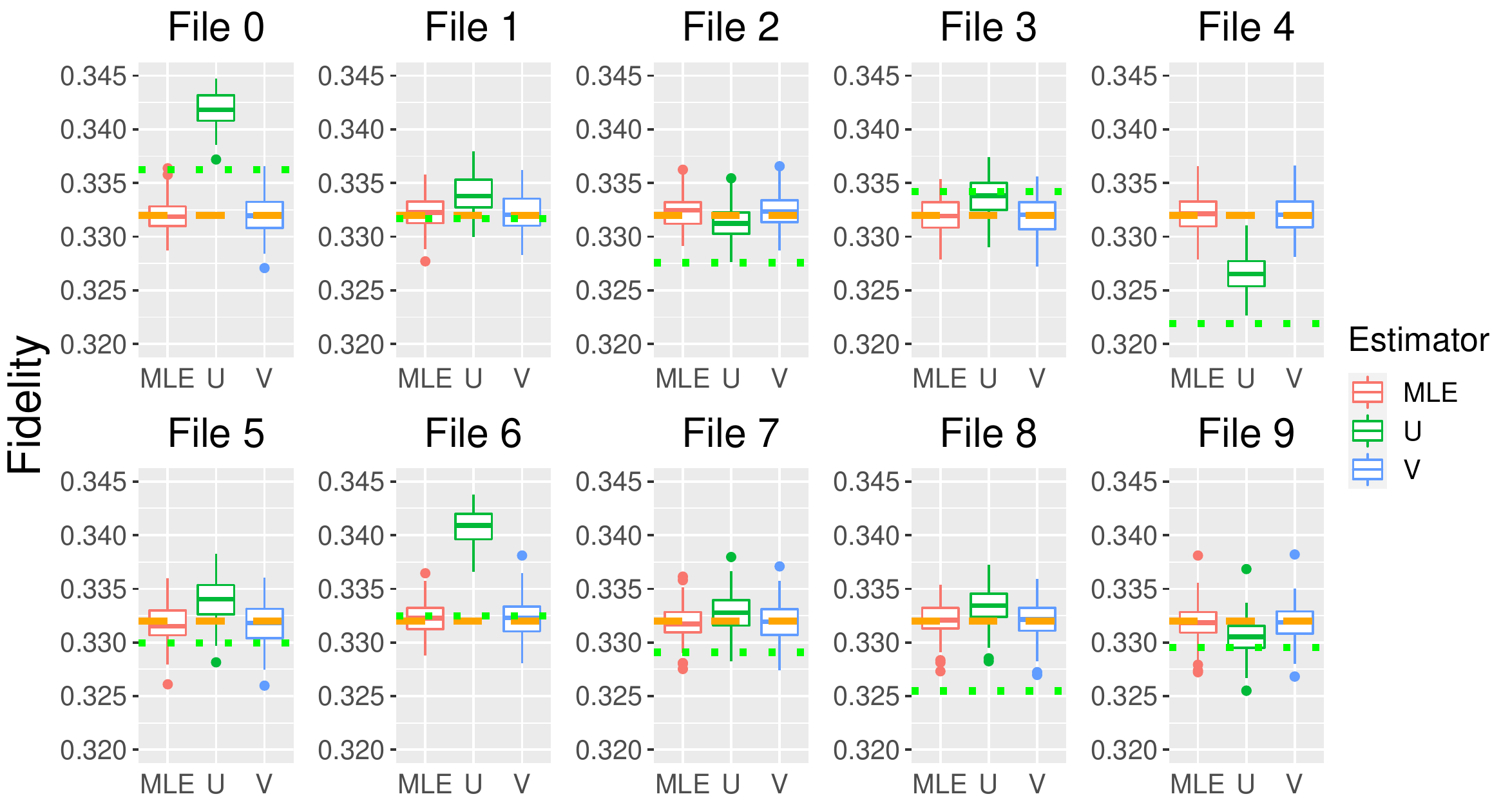}
		\caption{Boxplots of 100 values of MLE, $U$ and $V$,  with $\phi=0.3320$ and $n$=14.}
		\label{fig:databox}
	\end{center}
\end{figure}

In Figure \ref{fig:box}
we generated files $\{w_i\}_1^M$ with $M=2^{12},\, 2^{14}$, and $2^{16}$, from the Dirichlet distribution, and sampled from them using \eqref{eq:googsbsample} with  $\phi= 0.3862,\, 0.3320$, and $0.2828$, respectively (indicated by  orange dashed lines), and  $N=500,\!000$. The above  values of $\phi$ correspond to values computed in \cite{Google}  using a slightly modified version of \eqref{e:77}. We averaged the estimator over 10 files for each $M$, as done in \cite{Google}. This was repeated 100 times. We know that the MLE and $V$ are unbiased even without averaging, whereas $U$ is biased.  The boxplots below show that the average value of $U$ over 10 files has a small bias, but its variance is larger than that of the other two estimators. 
We next present comparisons of MLE, $U$, and $V$ conditional on $\{w_i\}_1^M$, that is, for given circuits. This is demonstrated on Google's data sets. 
In Figure \ref{fig:databox} we took 10 files $\{w_i\}_1^M$ of size $M=2^n=2^{14}$, which are obtained from Google's data.
These files of probabilities  $\{w_i\}_1^{M}$ will be used several times below. As before,  we sampled $N=500,\!000$ values from each set of the 10 probability files $\{w_i\}_1^{M}$ according to \eqref{eq:googsbsample} with $\phi=0.3320$ (indicated by the dashed orange line).   This was repeated 100 times for each of the 10 files. The dotted green line represents Google's estimate $U$ of the fidelity, which we computed for each file. The boxplots show the MLE, $U$, and $V$ for 100 values each.

We see that if the true fidelity is 0.3320 as suggested by Google, then their  estimate $U$ is off target in each case (but only by 5\% or less). There are several possible explanations: first, we know that $U$ is biased. For some of the files Google's estimator is within the range of our simulated $U$'s, but for others it is not, suggesting that further explanations are needed. Other potential explanations are that the theoretical number $\phi=0.3320$ is not exactly the actual $\phi$ that generated the sample, or that there are some other deviations from \eqref{eq:googsbsample} that affect the estimator $U$ (despite its robustness).  When we compute Google's $U$ we see that indeed for each of the 10 files it is biased, and as expected the bias can be upward or downward, depending on the particular $\{w_i\}_1^{M}$. On the other hand, $V$ and the MLE  are always centered around the value of $\phi$ that was used in the simulation, which is no surprise as $V$ is unbiased for each given set $\{w_i\}_1^{M}$, and the MLE is consistent and hence nearly on target when $N$ is large. 

\begin{figure}[t!]
	\begin{center}
		\includegraphics[width=13.2cm, height=6.2cm]{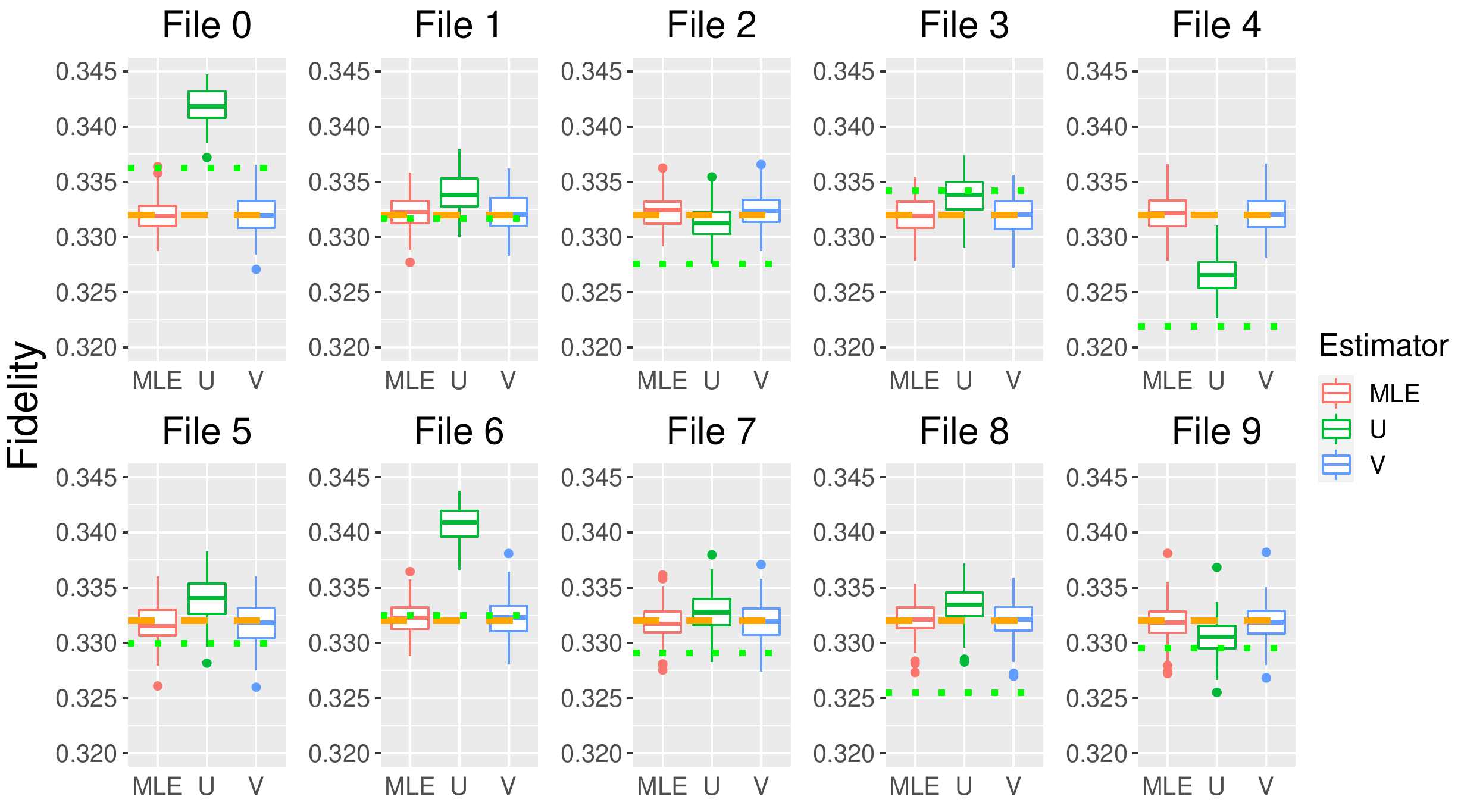}
		\caption{Boxplots of 100 values of MLE, $U$, and $V$, with $\phi$ = average MLE.}
		\label{fig:databoxMLE}
	\end{center}
\end{figure}

\begin{figure}[t!]
	\begin{center}
		\includegraphics[width=13.2cm, height=6cm]{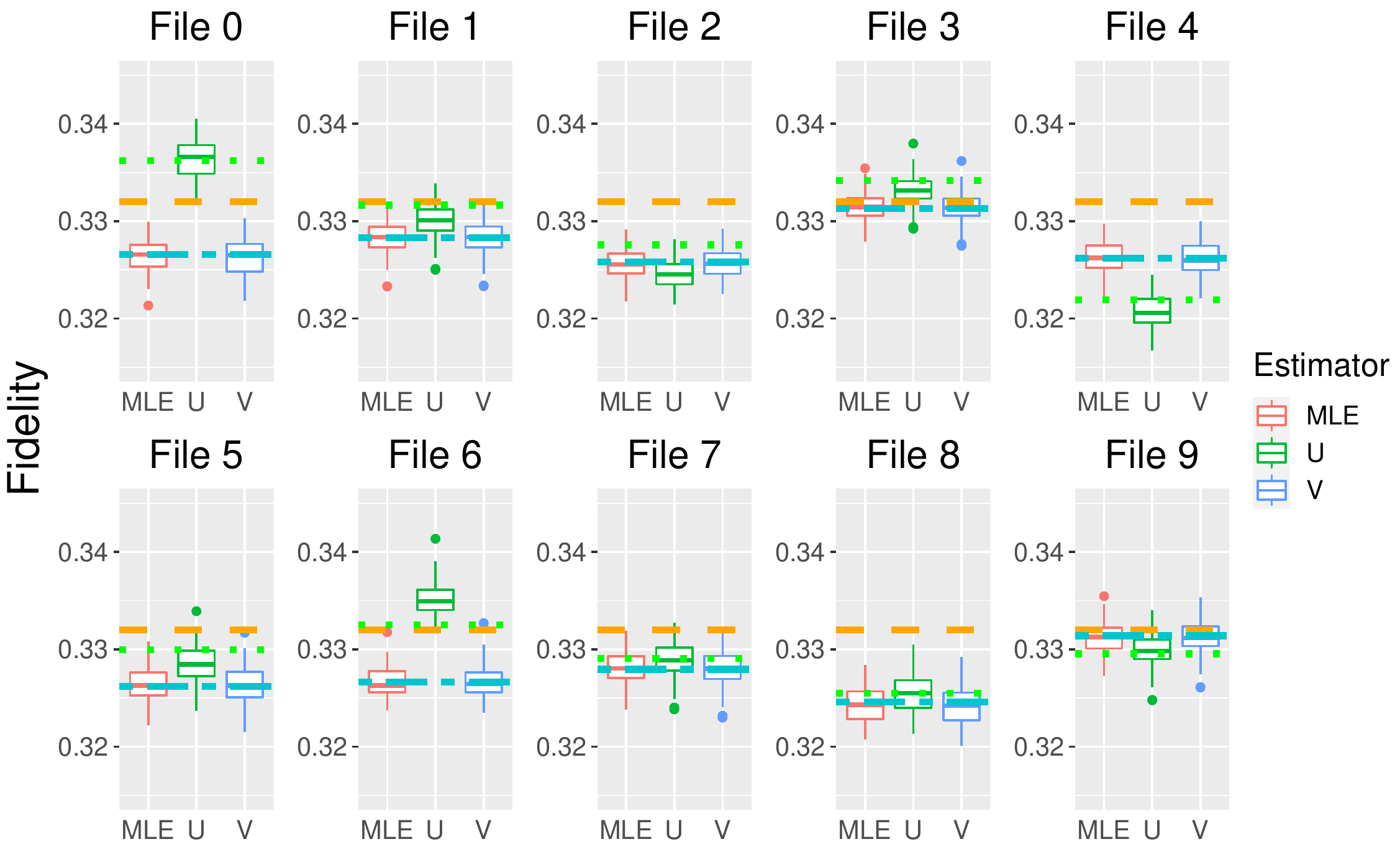}
		\caption{Boxplots of 100 values of  MLE, $U$, and $V$, with $\phi$=MLE of each file.}
		\label{fig:databoxMLE1}
	\end{center}
\end{figure}

Figure \ref{fig:databoxMLE} is similar to Figure \ref{fig:databox}, except that we sampled with the value of the average MLE over the 10 files  (indicated by the dashed turquoise line), rather than a theoretical $\phi$. In Figure \ref{fig:databoxMLE1} we calculated the MLE of each of the 10 files separately, and used it as $\phi$ (indicated by the dashed turquoise line) for the simulation based on \eqref{eq:googsbsample}. This is the most accurate way to simulate Google-like experiments based on the full list of probabilities $\{w_i\}_1^M$.
 Figure \ref{fig:databoxMLE1} suggests that the fidelity varies only slightly between circuits, and that the MLE
 represents the actual value of $\phi$ for each circuit. 


\subsection{Further analysis of Google's estimator}	

Although the MLE is in general better than $U$, it is of interest to consider explicit estimators like $U$.
We next show that  Google's choice of $U$  is   best in the sense of minimizing the variance  as $\phi \to 0$ in a class of linear estimators defined below,  while for larger values of $\phi$ another estimator, also discussed in \cite{Google}, is better. Some of  these results, which motivated their choice of $U$, are reported in the Supplement to \cite {Google}.  We have the following  proposition:

\begin{proposition}\label{prop:umvu}
	Assume that 
	$(w_1,\ldots,w_M) \sim$ Dirichlet$(\bf 1)$.
	Among all linear unbiased estimators of the form 
	$$U_g:=\frac{1}{N} \frac{\sum_{j=1}^N g(\widetilde w_j)-B}{A},$$ that is, with $A$ and $B$ such that $EU_g=\phi$,
	the choice $g(w)=w$ for which $U_g=U$ of \eqref{eq:linmestgoog} minimizes $Var(U_g)$ when  $M$ is large and $\phi =  0$, and in this sense $U$ is optimal as $\phi \to 0$.
	For large values of $\phi$ (approximately $\phi > 0.32$) 
	the estimator $U_g$ with $g(z)=\log(z)$ has a smaller variance than that of $U$.
\end{proposition}
\noindent{\it Proof:}
Let  $Y:=\frac{1}{N} \sum_j g(\widetilde w_j)$. 
We have
\begin{equation}\label{eq:EW}
E^w Y \equiv E(Y\mid \{w_{i}\})= E[g(\widetilde w_{j})\mid \{w_{i}\}] =\sum_{i=1}^Mg(w_{i})(\phi w_{i}+(1-\phi)/M).
\end{equation}
Unconditioning, that is, taking expectation in \eqref{eq:EW} over $w_i$, 
we have 
\begin{equation}\label{eq:omega}
EY=Eg(\widetilde w_{j})=M E[\phi g(w)(w-1/M)]+Eg(w)= \phi A +B,
\end{equation} 
where $A=ME[g(w)(w-1/M)]$, $B=Eg(w)$, and $w$ stands for a generic $w_i$.

We analyze the best approximate choice of $g$ for small  $\phi$ by  setting $\phi=0$. Since $U_g$ is unbiased, Equation \eqref{eq:pytha} implies that $Var(U_g)=E[Var(U_g) | \{w_i\}]$; 
we therefore consider $Var(Y)=E[Var(Y | \{w_i\})]$.
When $\phi=0$ we have uniform sampling and since conditionally on $\{w_i\}$ the $\widetilde w_i$'s are iid, we have that  $$Var[\sum_{j=1}^N g(\widetilde w_j) | \{w_i\}]= N Var[g(\widetilde w_j)| \{w_i\}]=\frac{N}{M}\sum_{i=1}^M [g(w_i)-\mu_g]^2$$
where $\mu_g=\sum_{i=1}^M g(w_i)/M$. It is easy to see that $\mu_g \to Eg(w_i)$ with probability 1 as $M \to \infty$ and therefore, $$E Var[\sum_{j=1}^N g(\widetilde w_j) | \{w_i\}]= E\frac{N}{M}\sum_{i=1}^M [g(w_i)-\mu_g]^2 = E{N}[g(w_i)-\mu_g]^2\to N Var[g(w_i)].$$
With $w$ replacing $w_i$ 
we  have $Var(w) =Ew^2-(Ew)^2 = 2/M(M+1)-1/M^2=(M-1)/M^2(M+1)$.  Recalling $A$ from \eqref{eq:omega}
we have for small $\phi$ and large $M$,
\begin{multline*} N Var(U_g) \approx \frac{NVar [g(w)]}{M^2\{E[g(w)(w-1/M)]\}^2}\\=\frac{N}{M^2 Var(w)}\frac{Var [g(w)]Var(w)}{\{E[g(w)(w-1/M)]\}^2}
=\frac{(M+1)}{(M-1)Corr^2(g(w),w)}.
\end{multline*}
Thus $Var(U_g)$ is minimized when the correlation takes its maximal value 1, which is equivalent to $g$ being linear.

We now analyze the choice $g(w)=\log(w)$, which is mentioned in the Supplement to \cite{Google} and seems natural from the
point of view of information theory and maximum likelihood theory.
Using the approximation $w_i=z_i/M$, and assuming $z_i \sim$ Exp(1) and some calculations given in the Appendix, we have 
\begin{equation}\label{eq:appoxxVrlog}
Var(U_{g(w)=log(w)}) \approx \frac{1}{N}(\pi^2/6-\phi^2) \approx \frac{1}{N}(1.6449-\phi^2).
\end{equation}
Equation \eqref{eq:varruu} with large $M$ implies 
\begin{equation}\label{eq:appoxxVr}
Var(U)  \approx \frac{1}{N}(2\phi-\phi^2+1).
\end{equation}
The expression in \eqref{eq:appoxxVr} is increasing in $\phi$ and that in 
\eqref{eq:appoxxVrlog} is decreasing. They intersect when $\phi= 0.6449/2=0.32$. It follows that for small values of $\phi$ Google's linear estimator of \eqref{eq:linmestgoog}, which coincides with $U_{g(z)=z}$, 
is better than $U_{f(z)=log(z)}$; however, for approximately  $\phi > 0.32$ we see that $U_{f(z)=log(z)}$ is better in the sense of having a smaller variance. 
\qed

The results given in \eqref{eq:appoxxVr} and \eqref{eq:appoxxVrlog}  and the   comparison between them
appear  in the Supplement to \cite{Google} (without proof and details).

\subsection{Further estimation of fidelity}\label{sec:TomerT}
As mentioned before, Google's estimator $U$ as well as the MLE require knowledge of  the probabilities $\{\widetilde w_{j}\}_1^N$ associated with the observed sample $\{{\bf x}_{{\bf \widetilde w}_j}\}_1^N$, whereas computation of $V$ requires knowledge of the whole set of probabilities $\{ w_{i}\}_1^M$, both considered hard to compute for large $M=2^n$.

We now propose an estimator of the fidelity $\phi$ that is a function only of the observed sample of bitstrings ${\cal S}_{\bf x}=\{\widetilde{\bf x}^{(j)}\}_1^N$ and, unlike the above-mentioned estimators, does not require any knowledge of the sampling probabilities. However, this estimator requires a large sample size $N$ relative to the circuit size $n$; therefore, it is proposed only for its conceptual value and not as a practical way to avoid the difficulty of computing the probabilities $\{ w_{i}\}_1^M$.

The estimator $T$ proposed next is closely related to the notion of speckle purity benchmarking (SPB) from the supplement to \cite{Google}, Section VI.C.4, and especially equation (49). This estimator was used in \cite{Google} for circuits of 2 qubits and was not computed for the experimental samples.

Given a sample ${\cal S}_{\bf x}$, define for any bitstring ${\bf x}^{(i)} \in \{0,1\}^n$
$$n_i:= \text{  number of bitstrings in   }  {\cal S}_{\bf x} \text{  that are equal to   } {\bf x}^{(i)}.$$
With 
$\pi_i:=\pi({\bf x}^{(i)})$ and sampling according to  \eqref{eq:googsbsample} we have 
$ n_i  \sim Bin(N,{\pi}_i)$.
In order to construct an estimator of $\phi$ we first compute 
\begin{multline}\label{eq:forTom}
E^w \Big[\sum_{i=1}^M n_i^2 \Big] = 
\sum_{i=1}^M E^w\left[n_i^2 \right] 
= \big(\sum_{i=1}^M  Var^w(n_i) +[E^w n_i]^2\big)\\
= \sum_{i=1}^M \left( N{\pi}_i(1-{\pi}_i) + N^2{\pi}_i^2\right)
= N +(N^2-N)\sum_{i=1}^M{\pi}_i^2.
\end{multline}
By \eqref{eq:googsbsample}, with $w^{(2)}:=\sum_{i=1}^M w_i^2$ as before, we have
\begin{equation*}
\sum_{i=1}^M{\pi}_i^2=\sum_{i=1}^M \big(\pi({\bf x}^{(i)})\big)^2 
= \sum_{i=1}^M \left(\phi w_i +{(1-\phi)}/{M}\right)^2 
= \phi^2(w^{(2)}-{1/M}) + {1}/{M}, 
\end{equation*}
and then by \eqref{eq:forTom}
$E^w \Big[\sum_{i=1}^M n_i^2 \Big]=N +(N^2-N)\big(\phi^2(w^{(2)}-{1/M}) + {1}/{M}\big).$
Recalling $Ew^{(2)}=2/(M+1)$ we readily obtain  by taking another expectation 
$$E\Big[\sum_{i=1}^M n_i^2 \Big] = N+(N^2-N)\big[\phi^2(M-1)/[M(M+1)]+1/M\big].$$ 

This implies that the  statistic  $$T^{[2]}:=\frac{M(M+1)}{(N^2-N)(M-1)}\left(\sum_{i=1}^M n_i^2 -N-(N^2-N)/M\right) $$
is an unbiased and consistent estimator of $\phi^2$.  It may take negative values, and we define  $T := \sqrt{\max\{T^{[2]},0\}}$, which  is again a consistent estimator of $\phi$.
Counting  negative values as zero causes an upward bias. On the other hand, taking the root  to estimate $\phi$ causes a downward bias that increases with the variance.

\begin{figure}[h]%Fig6
	\begin{center}
		\includegraphics[width=10cm, height=6cm]{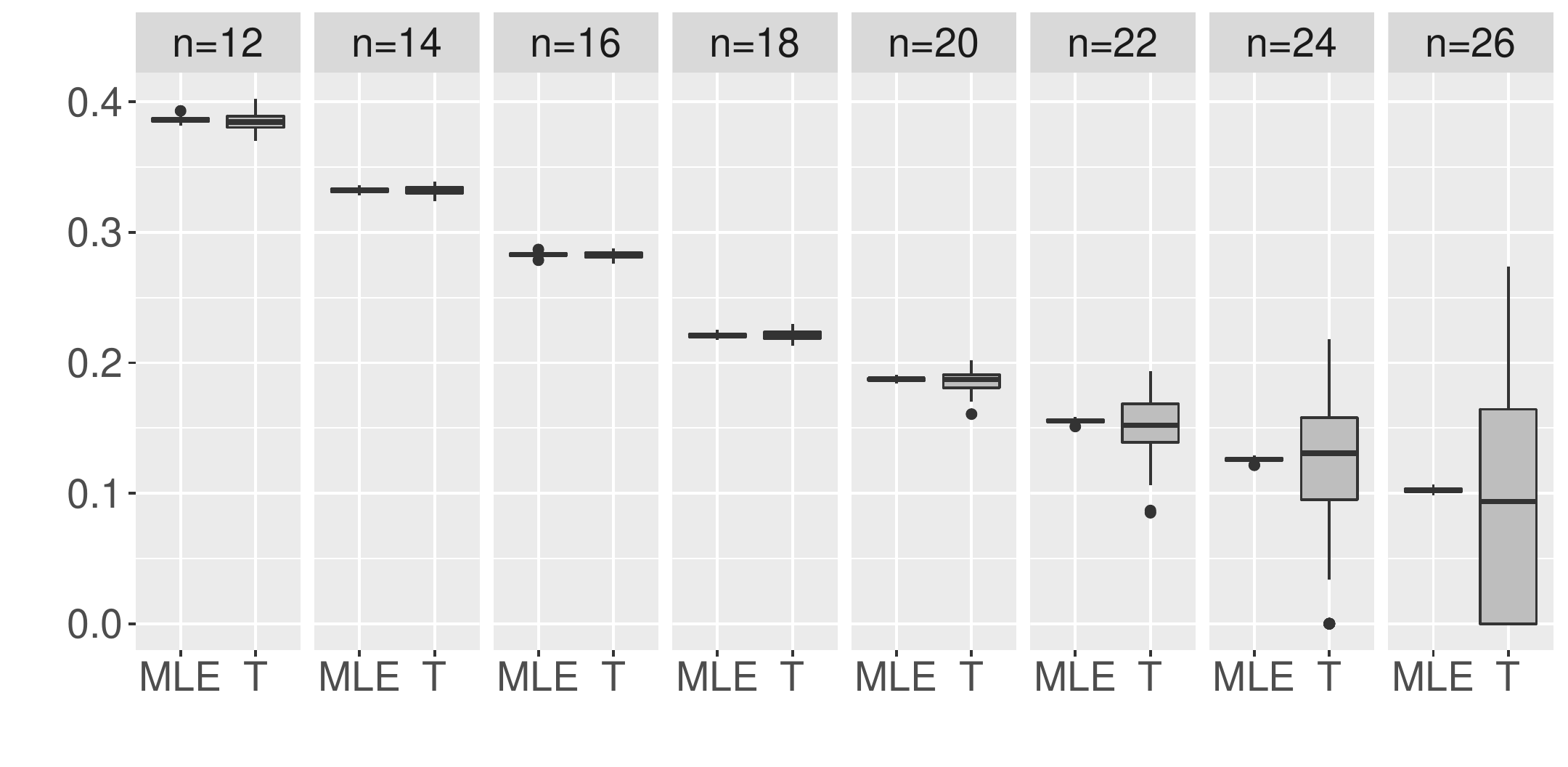}
		\caption{Boxplots of $T$ and the MLE}
		\label{fig:T}
	\end{center}
\end{figure}

Each boxplot in Figure \ref{fig:T} represents 100 simulated values of the statistic $T$ and the MLE  according to Google's model \eqref{eq:googsbsample}, which is an excellent estimator of the fidelity used for each simulation.  We simulated for circuits with even sizes  $n=12-26$ qubits, with the corresponding fidelities according to \eqref{e:77},  with sample size $N$=500,000.

We see that for $n=26$ the estimator $T$ has a very  large spread, and hence it becomes useless  for $N$ as above, which is of the order of Google's sample sizes.

Using Google's files, a comparison between average $T$ (of 10 files), average MLE, and the fidelity according to \eqref{e:77} (Equation (77) in the Supplement to \cite{Google}), given in the table below, shows that $T$ is biased upward.
It is easy to see that the statistic $T$ does not possess the robustness of $U$ described in Section \ref{sec:Rob}, and selection against part of the bitstrings will tilt $T$ upwards. Whatever the explanation of the overestimation of $\phi$ by $T$ in Google's file, it suggests that they deviate from the sampling model \eqref{eq:googsbsample}. We return to this issue in Section \ref{GS}.
\small{\begin{center}
	\begin{tabular}{||c c c c||} 
		\hline
		n&(77)&AVG MLE& AVG T \\ [0.5ex] 
		\hline\hline
		12&0.3862&0.3687&0.4689\\
		14&0.3320&0.3275&0.4392\\
		16&0.2828&0.2725&0.3917\\
		18&0.2207&0.2444&0.3557\\
		20&0.1875&0.2184&0.3210\\
		22&0.1554&0.1651&0.2989\\
		24&0.1256&0.1407&0.2838\\
		26&0.1024&0.1140&0.2600\\
		\hline
	\end{tabular}
\end{center}}

 Note that if the estimators $U$, $V$, and the MLE yield a significant value of $\phi$ for a  sample produced by a given circuit with given probabilities, it is evidence that the circuit performed a sampling task related to the given probabilities. This is not so for the estimator $T$, which cannot provide such evidence; its distribution does not depend on the particular circuit's probabilities $\{w_i\}$, and would be the same regardless of whether the circuit performed the sampling task using the claimed set of $\{w_i\}$, or another set generated at random according to the relevant Porter--Thomas distribution.
 
\section{The general case of $\lowercase{p}$}
\label{s:genp}
\subsection{A Google-type estimator}

We now briefly consider sampling according to \eqref{eq:sbsampleZ} for a general $p$. As mentioned before, this will be relevant when we consider more general sampling models that account for different kinds of sources of noise. The Google-type estimator of $\phi_k$ can be  defined as
\begin{equation}
\label{eq:linmestgoogg}
U_k=\frac{1}{N}\sum_{i=1}^N (M \widetilde w_{ki}-1).
\end{equation}
Using  the fact that $\phi_1+\ldots+ \phi_p=1$ and $\sum_{i=1}^M w_{ki}=1$, we have 
\begin{equation*}
E^wU_1=\sum_{i=1}^M (M w_{1i}-1)(\phi_1 w_{1i}+\ldots+ \phi_p  w_{pi})=M\phi_1w_{1}^{(2)}+M\sum_{k=2}^p \sum_{i=1}^M \phi_k w_{1i}w_{ki}-1,
\end{equation*}
where $w^{(2)}_{1}=\sum_{i=1}^M w_{1i}^2$.
For small $M$ the estimator is significantly  biased and like $U$ of \eqref{eq:linmestgoog}, the bias can be corrected if all $\{w_{ki}\}$ are known (in fact the sum of squares and inner products suffice). Taking another expectation we have
\begin{align*}
EU_1=EE^wU_1=E[M\phi_1w^{(2)}_{1}+M\sum_{k=2}^p \sum_{i=1}^M \phi_k w_{1i}w_{ki}]-1\\=2\phi_1 M/(M+1)-\phi_1=\phi_1\left(\frac{M-1}{M+1}\right),
\end{align*}
where we assume that $\{w_{ki}\}$ and  $\{w_{\ell i}\}$ are independent (or just uncorrelated) for all $k \neq \ell$  and satisfy the moment conditions in \eqref{eq:mom}. In fact, it suffices to assume that $Ew^{(2)}_{1}=2/(M+1)$ and $Ew_{ki}=1/M$.
Thus, $U_k$ is asymptotically  unbiased in $M$ when considered over random circuits.

\subsection{Maximum likelihood estimation}
\label {MLE}

The  log likelihood of the sample is
$$\ell({\boldsymbol \phi}) = \sum_{i=1}^N \log(\phi_1 \widetilde w_{1i}+\ldots+ \phi_p \widetilde w_{pi})=\sum_{i=1}^N\log[\phi_1 (\widetilde w_{1i}-\widetilde w_{pi})+\ldots+ \phi_{p-1}(\widetilde w_{(p-1)i}-\widetilde w_{pi})+\widetilde w_{pi}].$$
The maximum likelihood estimator 
of ${\boldsymbol \phi}= (\phi_1,\ldots,\phi_{p})$ such that $\phi_1+\ldots+\phi_{p}=1$ is obtained by computing the gradient ${\bf f}(\boldsymbol \phi)$ (also known as the score) and solving the equation 
\begin{equation*}\label{eq:ffzero}{\bf f}(\boldsymbol \phi)=\left(\frac{\partial}{\partial \phi_1 }\ell({\boldsymbol \phi}),\ldots, \frac{\partial}{\partial \phi_{p-1} }\ell({\boldsymbol \phi}) \right)=\bf 0,
\end{equation*}
where 
\begin{equation}\label{eq:partial}\frac{\partial}{\partial \phi_k }\ell({\boldsymbol \phi})=\sum_{i=1}^N\frac{(\widetilde w_{ki}-\widetilde w_{pi})}{\phi_1 (\widetilde w_{1i}-\widetilde w_{pi})+\ldots+ \phi_{p-1}(\widetilde w_{(p-1)i}-\widetilde w_{pi})+\widetilde w_{pi}}\,.
\end{equation}
%Lagrange multipliers with the constraint $\sum_{k=1}^p\phi_k=1$ yield the same equations, and again the solution needs to be projected to the $p$-simplex.

Starting with root-$N$ consistent estimators such as the moment estimators described above, we compute the MLE's by the Newton--Raphson algorithm. As mentioned before, asymptotically, one iteration suffices.
Specifically,
form the $(p-1) \times (p-1)$ Hessian matrix 
$J(\boldsymbol \phi)$ by taking another derivative in
\eqref{eq:partial}  (see \eqref{eq:Hess}) and then iterate by ${\boldsymbol \phi}_1 = {\boldsymbol \phi}_0 - {\bf f}({\boldsymbol \phi}_0) J^{-1}({\boldsymbol \phi}_0)$ (see \eqref{eq:NewtonRaph}).

\section{Refined noise models and readout errors}
\label{sec:diffmod}

Recall that in Google's notation  ${\cal P}({\bf x}^{(i)})=w_{i}$  
is the (random) probability of observing the bitstring ${\bf x}^{(i)}$ ($i=1,\dots, M$)
in the noiseless situation, namely, when $\phi = 1$.
The values $w_i$ depend on the quantum circuit $C$ and they are modeled to   
behave as independent Exp(1) random variables normalized by their sum (that is, as a Dirichlet distribution).   
We write ${\cal P}_C({\bf x}^{(i)})$ instead of ${\cal P}({\bf x}^{(i)})$, to indicate the dependence on the quantum circuit $C$.
In Google's basic noise model \eqref{eq:googsbsample} the
  sampling probabilities are $\phi {\cal P}_C({\bf x}^{(i)}) + (1-\phi)/M$.   

We now consider refined descriptions of the noise, replacing   \eqref{eq:googsbsample} by sampling  probabilities such as
%\begin {equation}
$\phi_1 {\cal P}_C({\bf x}^{(i)})+ \phi_2 {\cal N}_C({\bf x}^{(i)}) +\phi_3/M$,
%\end {equation}
where $\phi_1 +\phi_2 +\phi_3=1$, and  ${\cal N}_C (x)$ represents noise 
that depends on the circuit $C$.

In the next section we describe a
noise model of this kind, where ${\cal N}_C (x)$ describes the effect of readout errors when there are no
gate errors.  ${\cal N}_C$ can be seen as a secondary weak signal within the 
Google data that is largely independent of the primary signal ${\cal P}_C$. This refines
the analysis of errors, and allows us to offer independent estimation of the
fidelity by using the tools of Sections \ref{sec:Gphi} and \ref{s:genp}. 
Our analysis of the cases $n=12$ and $14$ indeed identifies the presence of this secondary signal and largely agrees with the fidelity estimation based on the primary signal ${\cal P}_C$. In Section \ref{GS} we show that Google's samples are far
from fitting the sampling model (1.1). The models discussed here improve the fit, but not by much.

\subsection{A readout noise model}
\label {RO}

We consider a refined version of Google's model that treats carefully the effect of readout errors.
(Readout errors are discussed in Section VI.D in the Supplement to \cite {Google}.) 
%takes into account the possibility that a string $x$ is sampled due to a readout error.
For this study we define the {\it total gate fidelity} $\phi_g$ as the probability that no errors occurr in the gate operations.
Two crucial ingredients of the Google experiment are the prediction of the fidelity
(\ref{e:77}) based on the fidelities of the
individual components, and the statistical estimator (\ref {google-estimator}) of the fidelity
based on the experimental bitstrings.
The  statistical assumptions that allow good predictions of the fidelity also
allow prediction of the total gate fidelity $\phi_g$,
and an extension of the Google estimator allows an estimation of $\phi_g$ from the experimental bitstrings.

We need to consider first how to estimate $\phi_g$ from Equation (\ref {e:77}), and then how to estimate
$\phi_g$ from the experimental bitstrings. 
The first task is easy. Equation (\ref {e:77}) (Google's Formula (77)) gives us an approximate description
of the total gate fidelity $\phi_g$ via its relation with the fidelity $\phi$:
$$\phi = \phi_g \cdot \prod_{q \in {\cal Q}}(1-e_q).$$
In this section we use a further simplification based on (\ref {e:77s}), namely, $\phi = \phi_g (1-0.038)^n,$
where $n$ is the number of qubits. Let $\phi_{ro}=\phi_g - \phi$ be the probability that there are
no gate errors but there are some readout errors.

We now discuss the estimation of $\phi_g$ based on the experimental bitstrings, a task that fits our general framework.
As before, the quantum circuit $C$ defines a probability ${\cal P}_C({\bf x}^{(i)})=w_i$, also denoted by $w_{{\bf x}^{(i)}}$, $i=1,\ldots,M =2^n$, where ${\bf x}^{(i)} \in \{0,1\}^n$.  
The Hamming distance between such strings is relevant to the nature
of noise considered in this section. 
Assuming that in the computation process there are no gate errors, 
we can observe a bitstring $\bf x$ in two ways:
\begin {enumerate}
\item 
Reading the output without error, and therefore
observing ${\bf x}$ with probability ${\cal P}_C({\bf x})=w_{{\bf x}}$, or
\item
Observing ${\bf x}$ because the true output
is ${\bf x}\oplus{\bf y}$ for some ${\bf y} \neq \bf 0$, an event whose probability
is ${\cal P}_C({\bf x}\oplus{\bf y})$ (where $\oplus$ denotes the XOR
operation, that is, mod-2 addition) and then readout errors occur exactly in the
coordinate $i$ in which $y_i=1$, an event whose probability is $q^{|{\bf y}|}(1-q)^{n-|{\bf y}|}$,
where $q$ is the probability of an individual readout error. 
\end {enumerate}
We assume independent errors with a common probability $q$, and so
the components of ${\bf y}$ are iid Bernoulli$(q)$ and the probability of ${\bf y}$ is
defined as ${\cal B}_q({\bf y}) := {\cal B}_q(|{\bf y}|)=q^{|{\bf y}|}(1-q)^{n-|{\bf y}|}$.
%A coordinate of $\bf y$ that takes the value one indicates a reading error. 
We can take $q=0.038$; see Equation (\ref{e:77s}) 
and \cite{Google}.
Let $D := P({\bf y}\neq {\bf 0})=\sum_{|{\bf y}| \neq 0} {\cal B}_q({\bf y})=1-(1-q)^n$,
and  let ${\cal N}_C^{ro}({\bf x})$ denote the probability of observing ${\bf x}$ due to
the second reason (point 2. above), conditioned on the existence of  readout errors only  (and no gate errors). Then  
\begin {equation}\label{eq:Nro}
{\cal N}_C^{ro}({\bf x})=\frac{1}{D}\sum_{{\bf y} \in {\{0,1\}^n},{\bf y} \neq {\bf 0}} {\cal P}_C({{\bf x}\oplus {\bf y}}){\cal B}_q({\bf y})= \frac{1}{D}\sum_{{\bf y} \in {\{0,1\}^n},{\bf y} \neq {\bf 0}} w_{{\bf x}\oplus {\bf y}}{\cal B}_q({\bf y}).
\end {equation}

We consider the noise model for the quantum circuit $C$ that produces a sample of size $N$
of $n$-strings , $\widetilde {\bf x}_1,\ldots \widetilde{\bf x}_N$, by drawing ${\bf x}^{(i)}$'s
independently $N$ times where the probability of drawing ${\bf x} \in \{0,1\}^n$ is
given by 
\begin{equation}\label{eq:detmodel}
\pi(\widetilde{\bf x}^{(j)}={\bf x}) = \phi {\cal P}_C({\bf x}) + 
\phi_{ro}\, {\cal N}_C^{ro}({\bf x})
+ (1 - \phi_g)/M,\,\,\,j=1,\ldots,N.
\end{equation}
This is a refinement of Google's basic model  \eqref{eq:googsbsample}, offering a more detailed description
of the sampling probability when there are only readout errors and no gate errors.
When there are gate errors, we still assume that the samples have the uniform distribution.
As explained already in \cite {Google} and in the previous sections,
such a noise model will not make a difference in estimating $\phi$, based on the main signal ${\cal P}_C(x)$),
but it will give us an opportunity for an independent estimation of
$\phi_{ro}$ (and hence also $\phi$ itself) based on the secondary signal ${\cal N}_C^{ro}({\bf x})$.
Note also that \eqref{eq:detmodel} is a special case of our general model \eqref{eq:sbsampleZ}
from the Introduction.

Using the notation  $v_i
:=\sum_{|{\bf y}| \neq {\bf 0}} w_{ {\bf x}^{(i)}\oplus {\bf y}}{\cal B}_q({\bf y})/D$ we have $\sum_{i=1}^M v_i=1$,
and recalling that ${\cal P}_C({\bf x}^{(i)})=w_{{\bf x}^{(i)}} = w_i$, it is easy to see that
Equation \eqref{eq:detmodel} is equivalent to sampling $N$ times (with replacement) with probabilities 
\begin{equation}\label{eq:detmodelz}
\pi({\bf x}^{(i)})=\phi w_i +
\phi_{ro} v_i
+ (1 - \phi_g)/M,\,\,\,i=1,\ldots,M.
\end{equation}
Note that $v_i$ contains all the $w_j$'s except for $w_i$. The variables $w_i$ and $w_j$, being coordinates of the Dirichlet distribution, are very weakly dependent when $M$ is large. Thus the components of \eqref{eq:detmodelz} are almost independent. In particular, the relevant fact here is that $E(w_iw_j)=1/[M(M+1)]$ for $i \neq j$ (see \eqref{eq:mom}),
which for large $M$ is very close  to $1/M^2$ that would obtain had they been independent. 

\subsection {Estimating the total gate fidelity}
Let $\widetilde {\bf x}^{(j)}$ denote an observation, $j=1,\ldots,N$,
corresponding to $(\widetilde w_j, \widetilde v_j)$ where $\widetilde v_j
=\sum_{|{\bf y}| \neq {\bf} 0} w_{\widetilde {\bf x}^{(j)}\oplus {\bf y}}{\cal B}_q({\bf y})/D$.
To estimate $\phi_g$ or $\phi_{ro}$,  define a Google-type estimator
$$W=\frac{M}{N}\sum_{j=1}^N \widetilde v_j\,-1,$$ 
and again let $\quad U=\frac{M}{N}\sum_{j=1}^N \widetilde w_j \,-1$.
Note that computing $W$  requires knowledge of all ${\cal P}_C({\bf x}^{(i)})=w_i$ and not just those in the sample. 
Taking expectation of $W$ with respect to the sampling probabilities of \eqref{eq:detmodelz} conditioned on $\{w_i\}_1^M$ and $\{v_i\}_1^M$ and using $\phi_g=\phi+\phi_{ro}$, we have
\begin{align}\label{eq:Eroest}
E^wW=M\phi \sum_{i=1}^Mv_iw_i+M\phi_{ro}\sum_{i=1}^Mv_i^2-(\phi+\phi_{ro}), \\
E^wU=M\phi \sum_{i=1}^M w^2_i+M\phi_{ro}\sum_{i=1}^Mw_iv_i -(\phi+\phi_{ro}). \nonumber
\end{align}
For given $\{w_i\}_1^M$ and $\{v_i\}_1^M$ it is easy to see that $W$ and $U$ are averages of iid variables, and hence they are consistent estimators of their expectations.  The method of moments (see Section \ref{ME}) applied here
consists of replacing $E^wW$ by the statistic $W$ and $E^wU$ by $U$ in  the above two equations and solving the resulting linear system for $\phi$ and $\phi_{ro}$.

We next consider $EW=EE^wW$. Define
$
G:= \frac{M}{M+1}\big\{[q^2+(1-q)^2]^n-2(1-q)^n+1\big\}, 
$
and recall that $D=1- (1-q)^n$.
In the  Appendix we show that
\begin{equation}\label{eq:EWRO2}
EW = \phi_{ro}[G/D^2-1]-\frac{1}{M+1}\phi\,,
\end{equation}
and neglecting the last term,   the estimator
\begin{equation}\label{eq:EstRO2}
\widetilde \phi_{ro}:=\frac{1}{G/D^2-1}W
\end{equation}
is nearly unbiased for $\phi_{ro}$ when expectation is taken with respect to both the sampling in \eqref{eq:detmodelz} and  with respect to all $\{w_i\}_1^M$ and $\{v_i\}_1^M$, that is, in the same sense that Google's $U$ is nearly unbiased (see \eqref{eq:EUUU}).
The average results for $U$ and $\widetilde \phi_{ro}$  for the ten files ($n$ = 12) are (0.3674, 0.1956). 

The computation of $Var(W)$ is rather cumbersome and will not be presented.
For data analysis purposes we shall compute it by simulations. 

Using $Ew_iv_i =E(w_iw_j)= 1/(M(M+1))$ for $i \neq j$, it is easy to see that $U$ is nearly  unbiased for $\phi$ since  $EU=\phi\frac{M-1}{M+1}-\phi_{ro}/(M+1)$. 

\subsection{MLE of $(\phi, \phi_{ro})$}
We rewrite the sampling rule \eqref{eq:detmodelz} as 
\begin{equation}\label{eq:detmodelzb}
\pi({\bf x}^{(i)})=\phi (w_i -1/M)+
\phi_{ro} (v_i -1/M)
+1/M,\,\,\,\,i=1,\ldots,M.
\end{equation}
The log-likelihood function is
$$\ell({\boldsymbol \phi}) = \sum_{j=1}^N \log[\phi (\widetilde w_j -1/M)+
\phi_{ro} (\widetilde v_j -1/M)
+1/M].$$
The maximum likelihood estimator 
of ${\boldsymbol \phi}= (\phi, \phi_{ro})$  is obtained by computing the gradient ${\bf f}(\boldsymbol \phi)$  and solving the equation ${\bf f}(\boldsymbol \phi)=\left(\frac{\partial}{\partial \phi }\ell({\boldsymbol \phi}), \frac{\partial}{\partial \phi_{ro} }\ell({\boldsymbol \phi}) \right)=\bf 0$,
where 
\begin{align*}\frac{\partial}{\partial \phi}\ell({\boldsymbol \phi})=&\sum_{j=1}^N\frac{(\widetilde w_j -1/M)}{\phi (\widetilde w_j -1/M)+
	\phi_{ro} (\widetilde v_j -1/M)
	+1/M}\quad \text{  and}\\ &\frac{\partial}{\partial \phi_{ro}}\ell({\boldsymbol \phi})=\sum_{i=1}^N\frac{(\widetilde v_j -1/M)}{\phi (\widetilde w_j -1/M)+
	\phi_{ro} (\widetilde v_j -1/M)
	+1/M}.
\end{align*}
Starting with root-$N$ consistent estimators such as the moment estimators described above, we
approach the MLE's by the Newton--Raphson algorithm. As mentioned before, asymptotically, one iteration suffices.
For the Newton--Raphson algorithm we need 
 the\, $2 \times 2$ \,Hessian matrix $J({\boldsymbol \phi})$, which is easy to compute by taking second derivatives of the derivatives given above.
The Newton--Raphson iteration can now be written as 
${\boldsymbol \phi}_1 = {\boldsymbol \phi}_0 - {\bf f}({\boldsymbol \phi}_0) J^{-1}({\boldsymbol \phi}_0).$

We estimated $\phi$ and $\phi_{ro}$ under the sampling model \eqref{eq:detmodel} -- \eqref{eq:detmodelz} for ten Google files for $n=12$ and sample size $N=500,\!000$. For these files formula \eqref{e:77} gives $\phi=0.3862$. The average of the MLE estimates of the pair $(\phi, \phi_{ro})$ over the ten files is $(0.3687, 0.1958)$.
The average MLE computed as in Section \ref{sec:condMLE} with sampling according to \eqref{eq:googsbsample}
comes out to the same value 0.3687, and individual estimates in the ten files differed slightly in the third or
fourth digit after the decimal point when estimated according to the two models.

\subsection {Related noise models}

We remark briefly on other noise models. In \eqref{eq:Nro} the random noises ${\cal N}_C^{ro}({\bf x})$
for different ${\bf x}$'s are dependent, with stronger dependence between ${\bf x}$'s that are close together in
the Hamming distance.
The noise term ${\cal N}_C^{ro}({\bf x})$ is a sum of $w_i$'s with weights. Using the representation $w_i=z_i/\sum_{j=1}^M z_j$ with iid $z_j$'s, it can be shown, e.g., by Lyapunov's central limit theorem, that for
fixed $\bf x$ the sum composing the error  ${\cal N}_C^{ro}({\bf x})$
 is approximately normal for large $M$.
A Gaussian assumption simplifies the dependence structure across ${\bf x}$'s and allows further calculations. 
One can also consider replacing the Binomial distribution 
${\cal B}_q({\bf y})$ in \eqref{eq:Nro} by another, for example,  a distribution that is supported only on $|{\bf y}| \le k$, thus
taking into account only neighbors of ${\bf x}$ having Hamming distance $\le k$, with, say, $k$ = 1 or 2 or so. This  approximation can
simplify the necessary computations. 

Finally, as it turns out, for the actual readout errors of the Google device the probability $q_1$ that 1 is read as 0 is 0.055 and the probability $q_2$ that 0 is read as 1 is  0.023.
This leads to the following sampling model:

\begin {equation}
\label {e:nqaro}
\pi ({\bf x})
= \phi_g \sum _{{\bf y} \in \{0,1\}^n} {\cal P}_C({{\bf x}\oplus {\bf y}}) q_1^{a}q_2^b (1-q_1)^c(1-q_2)^d~+~ (1-\phi_g){\cal B}_q({\bf x}),
\end {equation}
where $q_1=0.055$,  $q_2=0.023$,  
$a=|\{i: y_i =1 {~\rm and~} x_i=0\}|$,
$b=|\{i: y_i =1 {~\rm and~} x_i=1\}|$,
$c=|\{i: y_i=0 {~\rm and~} x_i=1\}|$, 
$d=|\{i: y_i=0 {~\rm and~} x_i=0\}|$,
and ${\cal B}_q({\bf x}) = {\cal B}_q(|{\bf x}|)=q^{|{\bf x}|}(1-q)^{n-|{\bf x}|}$, where now $q=(1-q_1+q_2)/2=0.484$. 

The first term of \eqref{e:nqaro} represents the probability that $\bf x$  is sampled, conditioned on no gate errors. 
Here ${\cal P}_C({{\bf x}\oplus {\bf y}})$ is the probability that without noise ${{\bf x}\oplus {\bf y}}$ will
be produced by the circuit. If, for example,  $x_i=0$ and $y_i=1$ (or $y_i=0$) then the $i$th coordinate  will
be read as 0 if a readout error occurs with probability $q_1$, corresponding
to $a$ above (if no readout error occurs, with probability $1-q_2$, corresponding to $d$).

We continue to assume that the effect of gate errors is to replace the desired distribution with a uniform probability but the asymmetric readout errors make a difference also there. %for the third term $(1/M)$ of \eqref{eq:detmodel}.
If uniformly distributed $n$-vectors are read with errors such that 1 is read as 0 with probability $q_1 = 0.055$ and 
0 is read as 1 with probability $q_2=0.023$ then their distribution is no longer uniform. Instead, for each coordinate the probability of 1 becomes   $(1-0.055)/2+ 0.023/2=0.484$,
and the probability of 0 is 0.516.

When $n=12$ we have by \eqref{e:77} $\phi=0.3862$ and then $\phi_g=\phi/(1-.038)^{12}=0.6148$.
Using ten Google samples of size $N=500,\!000$ with $n=12$ the average MLE of  $\phi_g$ is 0.5880. 
Using \eqref{e:nqaro} we computed the MLE with $(\phi_g, q_1, q_2)$ as parameters, 
and obtain the estimates $(0.5571, 0.0465, 0.0196)$ for Google's file 1 (with similar results
for the other files for $n=12$). Thus our estimate of $\phi_g$  and the value derived from \eqref{e:77}
are all within at most 10\% of each other, while our estimates of $q_i$ differ from Google's numbers by about 15\%.
The fact that the fidelity estimation based on \eqref{eq:detmodel} and \eqref{e:nqaro} largely agree can be seen as a demonstration of the robustness discussed in Section \ref{sec:Rob}. 
The model \eqref{e:nqaro} can further be improved by taking into account the individual readout errors for different qubits.

\section{Confidence intervals}\label{sec:ci}

In this section we compute confidence intervals for $\phi$ for the three estimators discussed in Section \ref{sec:estfi}. We start with $V=\left[\frac{1}{N}\sum_{j=1}^N
M\widetilde w_{j}-1\right]\Big/(M w^{(2)}-1)$ of \eqref{eq:GV},
which is unbiased. For the computation of  $V$,  knowledge of  the  sample values $\{\widetilde w_j\}_1^N$ is not enough, and it requires knowledge of $w^{(2)}$, which is a function of $\{w_i\}_1^M$. This differs from the computation of $U$ and the MLE, which are functions of  $\{\widetilde w_j\}_1^N$, and hence statistics. As mentioned before, this distinction is relevant if for some reason only the sample values $\{\widetilde w_j\}_1^N$ are available, and this holds for some of Google's data files.
The variance of $V$ conditional on the realization $\{w_i\}_1^M$ is given in \eqref{eq:varv}. We denote it now by \begin{multline*}\sigma_N^2(V;\phi,w^{(2)},w^{(3)})\\=
\frac{1}{N(M w^{(2)}-1)^2}\left[\phi(M^2  w^{(3)}-3M  w^{(2)}+2)-\phi^2(M  w^{(2)}-1)^2+M  w^{(2)}-1\right].
\end{multline*}
We assume that the realization $\{w_i\}_1^M$ is known, and therefore so are $w^{(2)}$ and $w^{(3)}$. If in a simulation $\phi$ is known, and we study the distribution of the estimator $V$, then we can use the known $\phi$ in computing $\sigma_N^2(V)=\sigma_N^2(V;\phi,w^{(2)},w^{(3)})$. If $\phi$ is unknown, it is standard practice to plug in an estimator, such as $V$ itself. Thus we assume that   $\sigma_N(V;\phi,w^{(2)},w^{(3)})$ is given. Since $V$ is based on a sum of iid variables it is asymptotically normal, and a standard 95\% confidence interval  is 
\begin{equation}
\label{eq:ciL1}\phi \in V \pm 1.96\, \sigma_N(V;\phi,w^{(2)},w^{(3)}).
\end{equation}
Both the normal and plug-in approximations suggest rounding 1.96 to 2; we shall keep 1.96 only because of its connotation. This is a \textit{conditional} (on the realization $\{w_i\}_1^M$) confidence interval.

Suppose now that we have $L$ independent samples, all taken under the same  $\phi$ according to \eqref{eq:googsbsample}. These samples may arise from a single realization  $\{w_i\}_1^M$ or from several different ones, all known. If we
have estimates $V_1,\ldots,V_L$ then a natural
approach is to first  estimate  $\phi$ by $\overline V=\frac{1}{L} \sum_{i=1}^L V_i$ and plug it in to obtain variance estimates $\sigma_{N_1}^2(V_1),\ldots,\sigma_{N_L}^2(V_L)$. We then replace $\overline V$ by the  variance-minimizing convex combination of the estimators
$\overline{\overline V}=\sum_{i=1}^L \frac{V_i}{\sigma_{N_i}^2(V_i)}\Big/\sum_{i=1}^L \frac{1}{\sigma_{N_i}^2(V_i)}$, whose variance is $\frac{1}{\sum_{i=1}^L 1/\sigma_{N_i}^2(V_i)}$. This process can be iterated, that is, plug in $\overline{\overline V}$ to reestimate the variances, etc.; however, we shall not pursue this. We obtain the conditional confidence interval  
\begin{equation}\label{eq: ciLV}
\phi \in \overline{\overline V} \pm 1.96\, \sqrt{\frac{1}{\sum_{i=1}^L 1/\sigma_{N_i}^2(V_i)}},
\end{equation}
where in $\sigma_{N_i}^2(V_i)=\sigma_{N_i}^2(V_i;\phi,w^{(2)}_i,w^{(3)}_i)$ we set $\phi=\overline{\overline V}$.

Consider now the maximum likelihood estimator MLE discussed in Section \ref{sec:condMLE}. To compute the MLE, as well as  $U$, only the sample values $\{\widetilde w_i\}_1^N$ are needed.  We can repeat all the above steps and obtain the same kind of confidence intervals as in  \eqref{eq:ciL1} and \eqref{eq: ciLV} by now replacing  $V$ by  the MLE and $\sigma^2_N(V;\phi,w^{(2)},w^{(3)})$ by 
$$\sigma^2_N(MLE;\phi,w^{(2)},w^{(3)})=1\Big/N\sum_{i=1}^M \frac{( w_{i}-1/M)^2}{\phi  w_{i} +(1-\phi)/M};$$
see \eqref{eq:Fisher}. Since Google's estimator $U$ is conditionally biased, we cannot apply to it the same procedure to obtain a conditional confidence interval. Unconditional confidence intervals are discussed next. 

We now consider the above estimators when both $\{\widetilde w_i\}_1^N$ and $\{ w_i\}_1^N$ are random, and we do not condition on them. We refer to intervals as \textit{unconditional}. This seems to be the approach taken by Google \cite{Google}. In this case  the estimators considered above are practically unbiased (for large $M$) and their (approximate) variances are given in \eqref{eq:varrvv} and \eqref{eq:varMLE} (where the approximate conditional and unconditional variances are the same as those discussed in Sections \ref{sec:Gphi} and  \ref{sec:condMLE}) and in
\eqref{eq:varruu}, where the variance reflects the conditional bias of $U$.

\begin{figure}[h]
	\begin{center}
		\includegraphics[width=12.5cm, height=5cm]{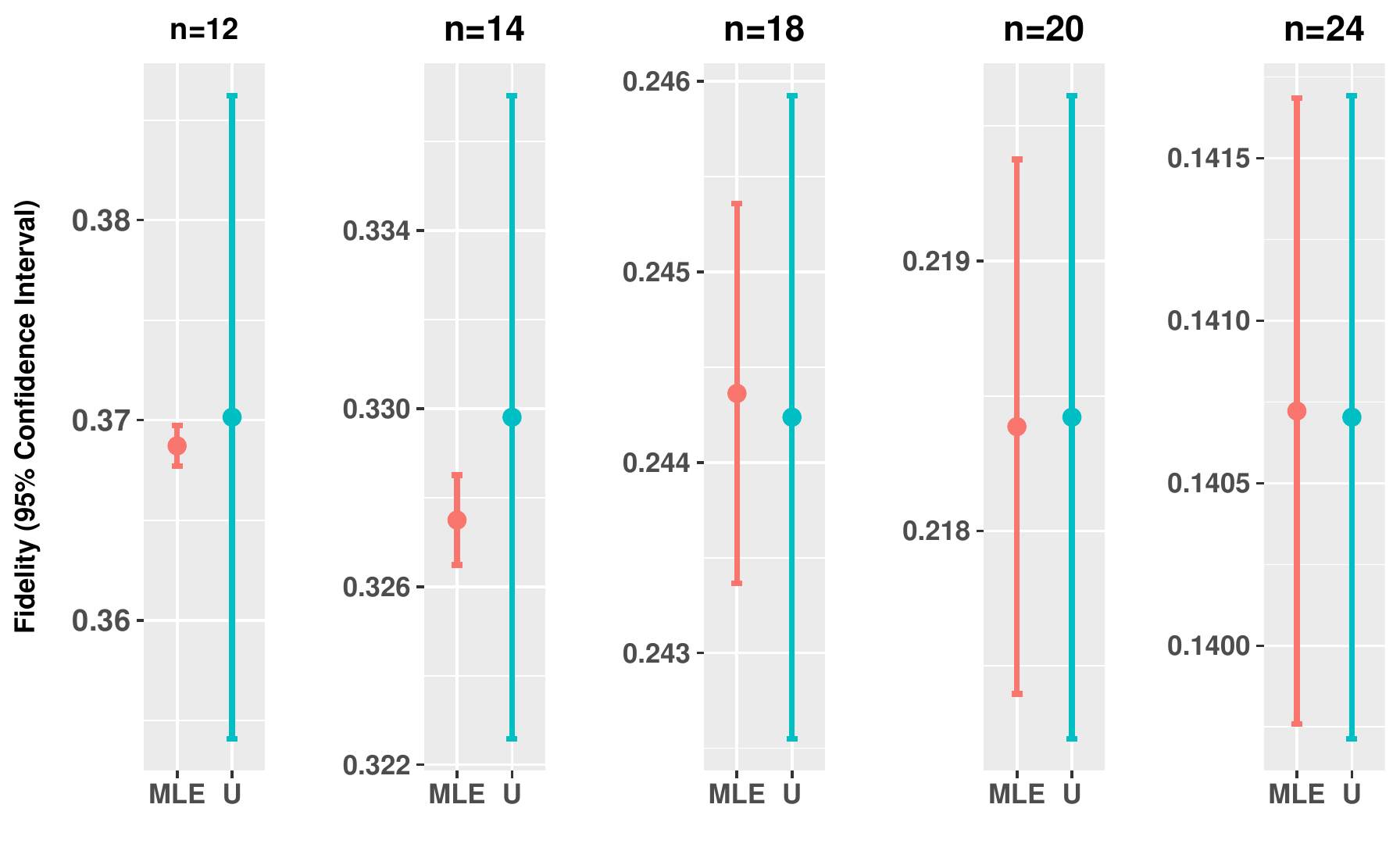}
		\caption{Unconditional confidence intervals for the fidelity based on 10 files for  circuits of different sizes $n$.}
		\label{fig:confint}
	\end{center}
\end{figure}

As above, we consider $L$ samples from $L$ different circuits with the same parameter  $\phi$, and $L$ estimators of $\phi$. In this case Google's  approach, in principle, is to average these estimators, leading to  the confidence interval
\begin{equation}
\label{eq:ciLU}
	\phi \in {\overline U} \pm 1.96\, \sqrt{ \frac{1}{LN}(2\phi - \phi^2+1)+\frac{20}{LM}\phi^2},
\end{equation}
where $\overline U=\frac{1}{L}\sum_{\ell=1}^L U_\ell$ is the average of the different estimators.
For the estimator $V$ we have in the same way 
\begin{equation}
\label{eq:ciLV}\phi \in {\overline V} \pm 1.96\, \sqrt{ \frac{1}{LN}(2\phi - \phi^2+1)},
\end{equation}
where $\overline V=\frac{1}{L}\sum_{\ell=1}^L V_\ell$, and 
for the MLE, we have by the results of Section \ref{sec:condMLE}
\begin{equation}
\label{eq:ciLMLE}\phi \in {\overline{MLE}} \pm 1.96\, \sqrt{ \frac{1}{LN\int_0^\infty \frac{(z-1)^2}{\phi z +1-\phi}e^{-z}dz}}.
\end{equation}
Figure \ref{fig:confint} shows confidence intervals computed using \eqref{eq:ciLU} and \eqref{eq:ciLMLE} with $L=10$ and $N=500,\!000$ for 10 Google files for each $n$. Note that the scale of the $y$-axis changes with $n$. We see that for $n=12$ and 14 the estimate $\overline U$ (the turquoise dot) is not in the confidence interval around the MLE, and the latter confidence interval is much smaller than the one around $\overline U$. As $n$ increases the estimates and their confidence intervals get closer together.

\section{Testing models and distributions}
\label {GS}

To demonstrate quantum supremacy, a convincing separation of the
fidelity estimator from zero is crucial.
But there are various reasons to try to understand
the actual state of the quantum computer from its experimental output  and its relation to probabilities  it represents.
There are also various ways, some mentioned in \cite {Google}, to compare the values
${\cal P}_C ({\bf x})$ for bitstrings obtained in an experiment with
the size-biased theoretical distribution, and this goes beyond the estimator of the fidelity. 
For small values of $n$, Google's sample sizes $N$ are rather large compared to $M=2^n$ and
we can test the goodness of fit of the
experimental output to the sampling probabilities.

Given a sample ${\cal S}_{\bf x}=\{\widetilde{\bf x}^{(j)}\}_1^N$,
with the sampling rule \eqref{eq:googsbsample}, let $n_i$ denote the number of sampled bitstrings that equal ${\bf x}^{(i)}$, that is,  $n_i = \big|\{j : \widetilde{\bf x}^{(j)} = {\bf x}^{(i)}\}\big|$, $i=1,\ldots,M$.
We have $\sum_{i=1}^M n_i = N$. As explained in Section \ref{sec:Int} there is an association between the
probabilities and bitstrings $ {\bf x}^{(i)}$ expressed by ${\cal P}({\bf x}^{(i)})=w_i=w_{{\bf x}^{(i)}}$.
Assuming all $\{w_i\}_1^M$ known, set $\pi_i = \phi w_i+(1-\phi)/M$, $i=1,\ldots,M$ according to the sampling model \eqref{eq:googsbsample}.  Pearson's chi-square test statistic is 
$${\cal X}^2=\sum_{i=1}^M\frac{(n_i-N\pi_i)^2}{N\pi_i},$$
and its asymptotic distribution for fixed $M$ and large $N$ is chi-square with $M-1$ degrees of freedom (or $M-2$ if $\phi$ is estimated). If the sample size $N$ does not suffice,
we can merge cells and probabilities, starting with cells with small $\pi_i$. 
The above test is relevant when we wish to verify the sampling model for a given realization $\{w_i\}_1^M$.
We looked at the 10 files given in \cite{Google} with $n=12$ and sample size $N=500,\!000$.

Using the $w_i$'s given in \cite{Google} and the MLE of the sample for $\phi$ we determine $\pi_i$ and
calculate ${\cal X}^2$ for Google's samples. We obtain very large ${\cal X}^2$ values,
of the order of 40,000. This is extremely significant and the $p$-value is practically zero
(since, for example, the level  $\alpha=10^{-15}$  critical values for $M-2$ = 4,094 degrees of freedom is about 4,855).  A rough calculation
shows that the average deviation of the cells' empirical probabilities from the theoretical $\pi_i$ is
about 0.25 standard deviations, and that the ${\cal X}^2$ statistic is
sufficiently large to reject the model with
about a tenth of the sample size used here,
and so the large ${\cal X}^2$ is not only due to the large sample size.
We computed the value of $\phi$ that minimizes the ${\cal X}^2$ statistic in the 10 Google files, and
find that it closely  agrees with the MLE; hence the large ${\cal X}^2$ is not  due to our
choice of $\phi$ in determining $\pi_i$. 
The  ${\cal X}^2$ statistics for the different readout models discussed in Section \ref{sec:diffmod}
were smaller by only about 10\%, and still extremely significant.   
The above test, the results on the estimator $T$ in
Section \ref{sec:TomerT}, and further results below, all indicate that the
sampling model \eqref{eq:googsbsample} does not provide an adequate description of the data.
Thus additional models should be explored.

\begin{figure}[]
	\begin{center}
		\includegraphics[width=12.5cm, height=8.7cm]{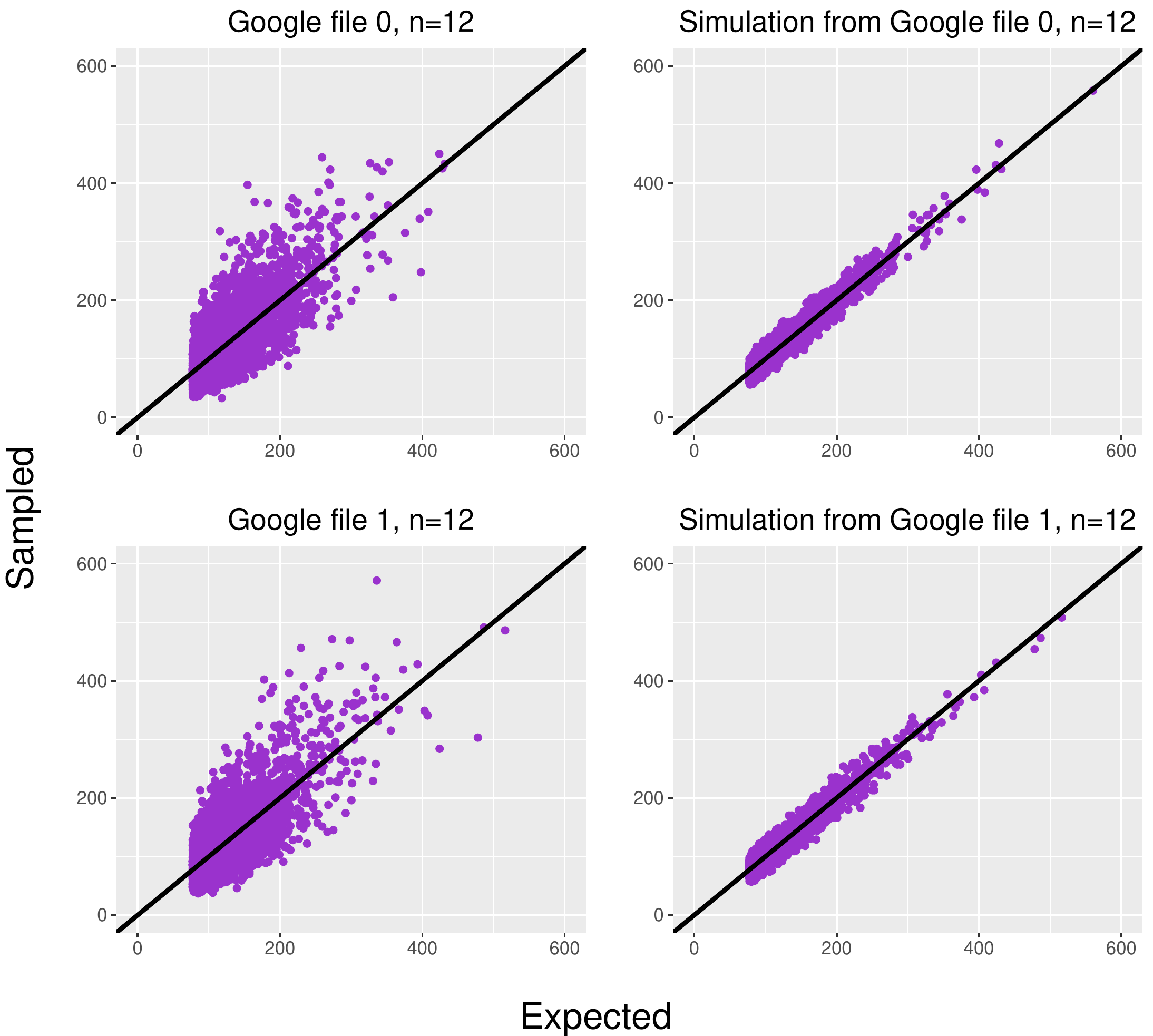}
		\caption{The left-hand side scatterplots display theoretical vs. empirical frequencies of two Google files with $n=12$. The
			right-hand side scatterplots display  theoretical vs. our simulated empirical frequencies
                       according to \eqref{eq:googsbsample}.}
		\label{fig:scattt}
	\end{center}
\end{figure}

In Figure \ref{fig:scattt} the two left-hand plots  show $N\pi_i$ against the corresponding empirical
frequencies for two files with $n=12$. These are compared (the two right-hand plots) to
similar plots using the same $\{w_i\}$'s for  a sample generated by \eqref{eq:googsbsample} with
$\phi = 0.3687$, which is the MLE estimated from the 10 files combined.
One can see
with the naked eye that Google's sample
is much more variable than a sample obtained according to the theoretical model. The truncation on the left is due to the fact that $N\pi_i  \ge N(1-\phi)/M=77$ in the present 
case.

\begin{figure}[]
	\begin{center}
		\includegraphics[width=13.5cm, height=3.6cm]{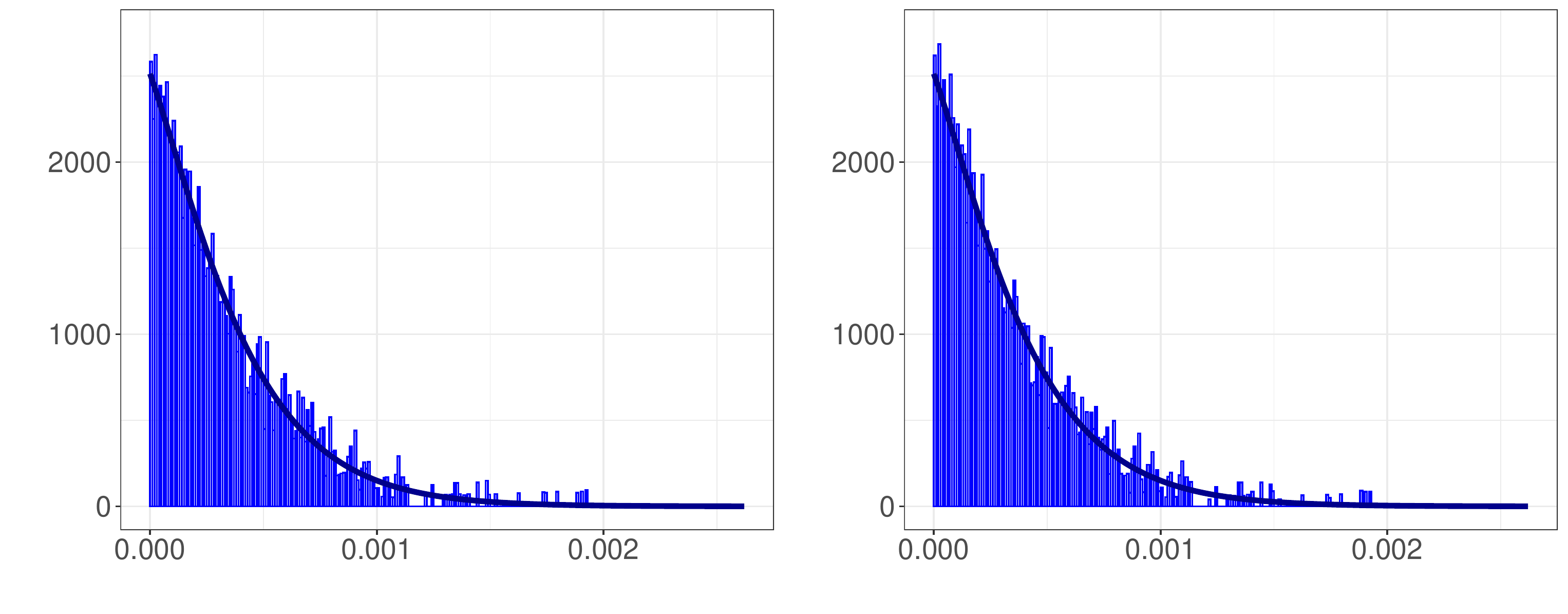}
		\caption{Histogram of a simulated sample $\{\widetilde w_i\}_1^N$ with $N=500,\!000$ from $\{ w_i\}_1^N$ of Google's file No. 0 $(n=12)$  according to \eqref{eq:googsbsample} with $\phi=0.3862$, 
			and histogram of Google's empirical frequencies. The black curves represent the asymptotic density of \eqref{eq:betamix}.}
		\label{fig:hists}
	\end{center}
\end{figure}

We plotted various histograms of Google's samples (for which the ${\cal X}^2$ values were huge)
and compared them to histograms of simulated samples based on the same $\{w_i\}_1^M$ and the MLE (for which ${\cal X}^2$ was
around 4,000 as expected).  Histograms involve smoothing and  with the naked eye 
we could not tell these histograms apart, in spite of the large difference in the ${\cal X}^2$ values.
Such histograms, with 200 cells each, are given in Figure \ref{fig:hists}.
The reader is invited to choose which histogram corresponds to Google's sample,
and which to our simulated sample using \eqref{eq:googsbsample}
with a non-significant ${\cal X}^2$ value.\footnote{On the right-hand side is Google's sample; on the left-hand side
  our simulated sample.}
To understand the theoretical distribution of such samples, recall that in Google's model $w_i=z_i/\sum_{j=1}^M z_j=\frac{z_i/M}{\sum_{j=1}^M z_j/M}\approx z_i/M$, where the approximation follows from the fact that $z_j \sim$ Exp(1) with mean = 1, and so the denominator $\to 1$ as $M \to \infty$.
If we approximate the sampling model \eqref{eq:googsbsample} by
$\pi({\bf x}^{(i)}) = \pi({ z}_i)=\phi {z_{i}}/M+(1-\phi)/M$ then, by Proposition \ref{prop{sizb}}, the
resulting sample $\{\widetilde {z}_{{\bf x}_i}\}$ will be approximately from
the distribution with density $\phi ze^{-z}+(1-\phi)e^{-z}$, which is a mixture
of Gamma(2,1) (the size-biased Exp(1)) and Exp(1) itself.
This approximation is used in \cite{Google}.

Similarly, samples obtained by \eqref{eq:googsbsample} from a Dirichlet vector $w_1, \ldots,w_M$ are
expected to have a distribution  that is close to the distribution whose density is a  mixture of Beta densities given by
\begin{equation}\label{eq:betamix} 
\phi M(M-1)t(1-t)^{M-2}+(1-\phi) (M-1)(1-t)^{M-2}.
\end{equation}
To see this note that
the marginal distribution of the Dirichlet vector is Beta(1, $M$-1) with density $(M-1)(1-t)^{M-2}$ for $t \in (0,1)$.
Therefore, the size-biased distribution has the Beta density  $M(M-1)t(1-t)^{M-2}$, and sampling 
by \eqref{eq:googsbsample} implies that the distribution of the sample 
$\widetilde w_1, \ldots,\widetilde w_N$  should be a discrete distribution (taking values from the set $\{w_i\}_1^M$) that is close to the above mixture by Proposition \ref{prop{sizb}}. This
is a mixture of Beta distributions, with the first part being the size-biased version
of the second (see Section \ref{sec:BSZ}).
From $\widetilde {z}_{{\bf x}_i} \sim \phi ze^{-z}+(1-\phi)e^{-z}$ we see that $\widetilde {z}_{{\bf x}_i}/M \sim \phi M^2 te^{-Mt}+(1-\phi)Me^{-Mt}$, which is seen to be approximately equal to \eqref{eq:betamix} by using $e^{-t}\approx1-t$.

The above analysis can be applied to more general noise models and we can consider size-biased distributions
not only with respect to the Dirichlet  or exponential distributions.
If a proposed noise model has several ingredients, e.g., 
$${\cal Q}({\bf x})=\phi_1 {\cal P}_C({\bf x})+ \phi_2 {\cal N}_2({\bf x})+\phi_3{\cal N}_3({\bf x})+\cdots,$$
then given a sample $\widetilde {\bf x}^{(1)},\dots, \widetilde {\bf x}^{(N)}$,
the statistical tools of the previous sections
can be used to estimate the different $\phi_i$'s, and we can test the distribution of
the sample as above for a fixed given realization or over different realizations.

It is important to note that  the fidelity estimators (apart from $T$ of Section \ref{sec:TomerT}) and the size-biased
distributions, when examined by standard histograms, are robust to various radical changes of the underlying probability distribution (see sections \ref{sec:wvsx} and \ref{sec:Rob}).
For example, they will not be affected if
we exclude from the sample all bitstrings with an odd number of 1's. They will also be robust to
various different noise models.
Such radical changes will greatly affect the empirical distribution and the outcomes of the chi-square test.

Returning to Figure \ref{fig:hists}, we note that the fact that the
estimator $T$ (Section \ref{sec:TomerT}) is considerably larger for the Google files compared to
the simulated samples generated by the theoretical models means that if we zoom into the individual cells,
we  see that
the samples generated by the quantum computers are more clustered  than the samples generated
by the theoretical models.

\section {Conclusion}\label{sec:conc}

We describe in statistical language the ideas and problems involved in quantum supremacy demonstrations, and study central
statistical aspects of the analysis of quantum computers' output involved in such demonstrations. The scientific and technological aspects of Google's demonstration are beyond the scope of this paper.

We study various related fidelity estimators.
Google's estimator $U$ of the fidelity, conditioned on a realization of the probabilities $\{w_i\}_1^M$, 
is biased in a way that depends on the realization. 
We study two unbiased estimators, MLE and $V$, that are more suitable for relatively small circuits.
As a fidelity estimator of an unknown realization, that is, when $\{w_i\}_1^M$ are considered random (with some assumptions),  the  estimator $U$ is unbiased
but has a larger variance compared to $V$ and the MLE. 
When the number of qubits increases and the fidelity decreases,  the gap between
these estimators diminishes and thus  
our results largely confirm the Google team's choice of the estimator $U$ for large-scale circuits.

Based on Google's readout analysis we considered various readout error models that
refine Google's basic model, and the associated estimators of the fidelity. A preliminary study
of the Google data on 12 and 14 qubits suggests that neither Google's basic noise model
nor our refined readout error model fits the observed data. 
An extension of the  analysis  beyond 12 and 14 qubits
can shed light on the statistical assumptions behind the Google fidelity
predictions and on other aspects of their experiment.
We note that there are further statistical aspects
of the experiments described in \cite {Google}, mainly those related
to the quality of individual components of the
circuit and to the process of ``calibration,'' which are not considered here.  

Finally, we expect that our study can contribute to the understanding of the nature of noise in noisy intermediate-scale
quantum computers.
Sampling tasks are natural candidates for testing and proving the potential
of quantum computers. Therefore,  our statistical analysis is
expected to be useful for various near-term experimental efforts for NISQ computers and 
future attempts for ``quantum supremacy" demonstrations.\\

\noindent\textbf{Acknowledgment} The numerical work in this paper was
done by the R Package and Wolfram Mathematica.
We are grateful to Larry Goldstein for insightful comments and questions, to Alexander Vlasov for helpful comments
on the Google data and other matters, and to  Carsten Voelkmann for two very careful readings of the manuscript and many invaluable  corrections. We are also thankful to members of the Google team
and especially to John Martinis and Sergio Boixo for helpful discussions. Two referees, an associate editor, and the Editor provided very useful comments on the content and organization of the paper.

\section{Appendix: Some proofs}

\noindent\textit{Proof of \eqref{eq:appoxxVrlog}, Proposition }\ref{prop:umvu}:  In order to compute $Var(U_{g(w)=\log(w)})$ we use the approximation (valid for large $M$) $w_i=z_i/M$, and we assume $z_i \sim$ Exp(1). We need the following facts:
\begin{multline}\label{eq:wolfram}\int_0^\infty \log(z)(z-1)e^{-z}dz=1,\,\,
\int_0^\infty \log(z)e^{-z}dz=-\gamma \approx -0.5772, \\
\int_0^\infty (\log(z)+\gamma)^2(z-1)e^{-z}dz=0\,\, and \,\,
\int_0^\infty (\log(z)+\gamma)^2e^{-z}dz = \pi^2/6 \approx 1.6449.
\end{multline} 
where $\gamma$ is  Euler's  constant.
The first equation in  \eqref{eq:wolfram} shows that in this case $A=1$; see \eqref{eq:omega}.
The second equation, with $w=z/M$ implies
that  $B=-\gamma-\log(M)$, and $EY=\phi-\gamma-\log(M)$, where $Y$ is defined in the beginning of the proof of Proposition \ref{prop:umvu} with $g(w)-\log(w)$. Thus, using the approximation $w_i=z_i/M$ in the third equality below,
\begin{align*} &Var[\log(\widetilde w_{j})] = EE^w[(\log(\widetilde w_j)-EY)^2\mid \{w_i\}] \\&=
E\{\sum_{i=1}^M(\log(w_i)-EY)^2[\phi w_i+(1-\phi)/M]\}\\
& \approx E \{(\log(z/M)-EY)^2[\phi z +(1-\phi)]\}
=\int_0^\infty (\log(z)-\phi+\gamma)^2[\phi(z-1)+1]e^{-z}dz,
\end{align*} and the above integrals imply
\begin{multline*}%\label{eq:appoxxVrlog}
\int_0^\infty (\log(z)-\phi+\gamma)^2[\phi(z-1)+1]e^{-z}dz\\=\int_0^\infty [(\log(z)+\gamma)^2+\phi^2-2\phi(\log(z)+\gamma)]
[\phi(z-1)+1]e^{-z}dz\\
=\int_0^\infty [(\log(z)+\gamma)^2]e^{-z}dz+\phi^2-2\phi^2+
2\phi\gamma-2\phi\gamma = \pi^2/6-\phi^2.
\end{multline*}
With $A$ and $B$ calculated above, we get $U_{g(z)=\log(z)}=\frac{1}{N} {\sum_{i=1}^N \log(\widetilde w_i)+\gamma+\log(M)}$ and
$Var(U_{f(z)=\log(z)}) \approx \frac{1}{N}(\pi^2/6-\phi^2) \approx \frac{1}{N}(1.6449-\phi^2)$. \qed\\

\noindent\textit{Proof of \eqref{eq:EWRO2}}\\
We claim that the expectation $EW$ taken with respect to the sampling 
rule \eqref{eq:detmodel} and then over  $\{w_i\}_1^M$ can be expressed after some calculations as  
\begin{align}\label{eqEEEWW} EW=\frac{M}{M+1}\phi+
M^2\phi_{ro}E\Big[\Big(\sum_{|{\bf y}| \neq 0} w_{{\bf x}\oplus {\bf y}}{\cal B}_q({\bf y})/D\Big)^2\Big]+(1-\phi_g)-1\nonumber   \\
=M^2\phi_{ro}E\Big[\Big(\sum_{|{\bf y}| \neq 0} w_{{\bf x}\oplus {\bf y}}{\cal B}_q({\bf y})/D\Big)^2\Big]-\phi_{ro}-\frac{1}{M+1}\phi\,,
\end{align}
where  $D := \sum_{|{\bf y}| \neq 0} {\cal B}_q({\bf y})=1- (1-q)^n$.
To see this take expectation over 
$E^wW$ in \eqref{eq:Eroest} whose  $M$ terms all have the same expectation; therefore, we can multiply by $M$ instead of summing. 
The  term $\phi$ in \eqref{eqEEEWW}  arises  from $$M\phi \sum_{i=1}^M E(v_i w_i)=M^2\phi E[\sum_{|{\mathbf y}| \neq {\bf} 0}w_{{\bf x}^{(i)}} w_{ {\bf x}^{(i)}\oplus {\bf y}}{\cal B}_q({\bf y})/D]=\frac{M}{M+1}\phi\,,$$ since $E(w_{{\bf x}^{(i)}}w_{ {\bf x}^{(i)}\oplus {\bf y}})=E(w_iw_j)=1/(M(M+1))$ by \eqref{eq:mom} (with $i \neq j$)  for $|{\bf y}| \neq 0$.
In order to compute $M^2Ev_i^2$ we compute 
$G:=M^2E\Big[\Big(\sum_{|{\bf y}| \neq 0} w_{{\bf x}\oplus {\bf y}}{\cal B}_q({\bf y})\Big)^2\Big]$
for any fixed ${\bf x}$. 
For off-diagonal terms in the square of the sum, that is, for  ${\bf y}' \neq {\bf y}$, we obtain again terms of the form $Ew_{{\bf x}\oplus {\bf y}}w_{{\bf x}\oplus {\bf y}'}=1/(M(M+1))$.
The total sum of these off-diagonal terms in $G$
is $\sum_{{\bf y} \neq {\bf y}', 1\leq |{\bf y}|, |{\bf y}'| \le n}{\cal B}_q({\bf y}){\cal B}_q({\bf y}')/(M(M+1))$.
For the diagonal terms the expectations satisfy $Ew_{{\bf x}\oplus {\bf y}}w_{{\bf x}\oplus {\bf y}}=2/M(M+1)$. For these terms we have 
$$H:=\sum_{|{\bf y}| \neq 0} {\cal B}_q({\bf y})^2=\sum_{k=1}^n {n \choose k}q^{2k}(1-q)^{2n-2k}=[q^2+(1-q)^2]^n-(1-q)^{2n},$$
and therefore 
$$K:=\sum_{{\bf y} \neq {\bf y}', 1\leq |{\bf y}|, |{\bf y}'| \le n}{\cal B}_q({\bf y}){\cal B}_q({\bf y}')=[1-(1-q)^n]^2-\{[q^2+(1-q)^2]^n-(1-q)^{2n}\}. $$
Therefore, 
\begin{multline}
\label{eq:Cread}
G=\frac{M}{M+1}
[{2H+K}] =\frac{M}{M+1}\big\{[1-(1-q)^n]^2+[q^2+(1-q)^2]^n-(1-q)^{2n}\big\}\\
=\frac{M}{M+1}\big\{[q^2+(1-q)^2]^n-2(1-q)^n+1\big\}. 
\end{multline}

It follows that 
\begin{equation*}\label{eq:EWRO2finn}
EW = \phi_{ro}[G/D^2-1]-\frac{1}{M+1}\phi\,,
\end{equation*}
which is \eqref{eq:EWRO2}. \qed

%\section{Appendix 2: Further data analysis}

\end {document}